\documentclass[showpacs,pra,twocolumn]{revtex4}%
\usepackage{amssymb}
\usepackage{amsfonts}
\usepackage{amsmath}
\usepackage{subfigure}
\usepackage{graphicx}
\usepackage[dvips]{color}%
\providecommand{\U}[1]{\protect\rule{.1in}{.1in}}

\begin{document}
\title{Finite--Temperature Density--Functional Theory of
Bose--Einstein Condensates} \author{Nathan Argaman}
\affiliation{Physics Department, NRCN, P.O. Box 9001, Beer-Sheva
84190, ISRAEL} \author{Y. B. Band} 
\affiliation{Departments of Chemistry and Electro-Optics and the Ilse
Katz Center for Nano-Science, \\
Ben-Gurion University of the Negev, Beer-Sheva 84105,
ISRAEL} 
\keywords{Density Functional Theory, Bose--Einstein Condensates} 
\pacs{03.75.Hh, 67.85.-d, 71.35.Lk} 

\begin{abstract}
The thermodynamic approach to density functional theory (DFT) is used to
derive a versatile theoretical framework for the treatment of
finite--temperature (and in the limit, zero temperature) Bose--Einstein
condensates (BECs). The simplest application of this framework, using the
overall density of bosons alone, would yield the DFT of Nunes (1999). It is
argued that a significant improvement in accuracy may be obtained by using
additional density fields: the condensate amplitude and the anomalous density.
Thus, two advanced schemes are suggested, one corresponding to a generalized
two--fluid model of condensate systems, and another scheme which explicitly
accounts for anomalous density contributions and anomalous effective
potentials. The latter reduces to the Hartree--Fock--Bogoliubov approach in
the limit of weak interactions. For stronger interactions, a local density
approximation is suggested, but its implementation requires accurate data for
the thermodynamic properties of \textit{uniform} interacting BEC systems,
including fictitious perturbed states of such systems. Provided that such data
becomes available, e.g., from quantum Monte Carlo computation, DFT can be used
to obtain high--accuracy theoretical results for the equilibrium states of
BECs of various geometries and external potentials.

\end{abstract}\date{\today}
\maketitle

\section{Introduction}

Our understanding of quantum--degenerate dilute Bose gas systems has
increased dramatically since the experimental achievement of
Bose--Einstein condensation (BEC) in ultra--cold dilute alkali gases
in 1995 \cite{Anderson,Bradley,Davis}.  The explosive growth of
knowledge of these systems has enriched both
atomic--molecular--optical physics and many--body physics.  Landmark
developments include the realization of control over the interaction
strength through magnetic--field tuning of a Feshbach resonance
\cite{Inouye,Cornish}, the generalization to Fermion systems, where
the crossover between BCS--type and BEC superfluidity has been
observed \cite{Regal}, nonlinear atom optics \cite{Deng_99},
mixed--phase condensates with different hyperfine states
\cite{Tripp00,Band01}, mixtures of different bosonic atoms (e.g.,
Na--Rb) \cite{Albus0211060}, mixed Bose--Fermi ultra--cold dilute gas
systems \cite{Maddaloni,Hu}, and studies where optical potentials have
been imposed on condensate systems, including systems in optical
lattices which are analogous to condensed--matter systems with either
weak or strong correlations \cite{Bloch_08}.  In the present work,
advanced methods for evaluating the finite temperature equilibrium
properties of a single--component dilute system of degenerate bosonic
atoms in an external potential will be studied.

A dilute gas of atoms behaves classically as long as the thermal de
Broglie wavelength, $\lambda_{T}=\sqrt{2\pi\hbar^{2}/mk_{\text{B}}T}$,
is smaller than the mean spacing between atoms, $n^{-1/3}$, where $T$
is the temperature, $n$ is the gas density, and $m$ is the atomic
mass.  Quantum degeneracy for a gas of bosonic atoms ensues when the
atomic wave packets begin to overlap, i.e., when $\lambda_{T} \gtrsim
n^{-1/3}$, and a condensate becomes populated.  For a uniform
noninteracting bosonic system, the critical temperature $T_{c}$, is
given by the condition $n\lambda_{T}^{3}=\zeta(3/2)\simeq2.612$, where
$\zeta$ is the Riemann zeta--function.  \ As the temperature is
lowered further, the kinetic energy of a uniform noninteracting gas
decreases, and vanishes in the $T\rightarrow0$ limit.  \ For a trapped
noninteracting gas, all the bosons occupy the ground state in this
limit, with the zero--point kinetic energy density balancing that of
the external potential.

In realistic Bose--condensed systems, interactions cannot be ignored,
as the interaction energy density is typically not small compared with
the kinetic--energy density (or with that of the external potential).
For systems which are dilute enough for the mean spacing between atoms
to be much larger than the range of the atomic potential \cite{vdw},
the interactions can be well modeled with a contact interaction with
an effective coupling constant $g=4\pi\hbar^{2}a_{0}/m$, which is
proportional to the two--body $s$-wave scattering length, $a_{0}$ (see
Ref.  \cite{Dalfovo}).  At zero temperature, the strength of the
interactions is quantified by the \textquotedblleft gas
parameter\textquotedblright, $na_{0}^{3}$.  In many practical
applications this parameter is very small, because $a_{0}$ is of
nanometer scale whereas the density of the alkali atoms at the moment
of condensation is of order several atoms per cubic micrometer.  A
reasonably good description of such systems can be obtained by using
the first--order expression for the interaction energy.  For high
accuracy, terms beyond the first order should nevertheless be taken
into account, because the density at the center of the trap increases
dramatically as the system is cooled below the condensation
temperature, and because the relative magnitude of the leading
correction is of order $10\sqrt{na_{0}^{3}}$ rather than of order
$na_{0}^{3}$.  Furthermore, the value of $a_{0}$ can be made much
larger near a Feshbach resonance, driving the system into a strongly
interacting regime.  At finite temperatures, the effects of the
interactions are more involved, as the ratio $gn/k_{\text{B}}T$ is
also an appropriate measure of the strength of the interactions, in
addition to the gas parameter.  If the gas parameter is small, this
ratio is also small at the transition temperature, but it can become
arbitrarily large upon decreasing the temperature.

In this paper a finite--temperature density functional theory (DFT) approach
to treat degenerate bosonic gases is developed. It is well known that DFT
provides both a rigorous conceptual framework and a set of highly--accurate
practical tools for calculating the ground--state properties of interacting
electron systems (for an introduction to DFT see
Refs.~\cite{Kohn99,Argaman,Capelle02}). Most calculations of the electronic
structure of atoms, molecules and solids are today carried out using the
Hohenberg--Kohn--Sham DFT approach introduced in the 1960's \cite{HK64,KS65}.
DFT has been generalized in many ways, e.g., to treat systems at
finite--temperature \cite{Mermin65}, in time--dependent external
fields,\cite{Bartolotti,Runge,Vignale} superconducting electronic
systems,\cite{Oliveira88} and systems as diverse as nuclei \cite{nuclei},
classical fluids \cite{Rowlinson}, spin density waves \cite{Capelle} and
superfluid liquid He \cite{HM65,DHPT90}. Another development in DFT is the
suggestion of using the principles of equilibrium thermodynamics to establish
the finite--temperature version of DFT as a fundamental thermodynamic
representation of the free energy, and viewing the ground--state DFT as the $T
\rightarrow0$ limit of this representation \cite{Argaman, AMspin}. Here, we
follow this approach, and apply it to dilute--gas bosonic systems.

DFT is a method for calculating the energy and density distribution of
an inhomogeneous system.  Within the Kohn--Sham approach, it employs a
noninteracting reference system which has the same density
distribution $n(\mathbf{r})$ as the fully interacting system.  This
noninteracting system is associated with an effective potential which
is distinct from the external potential of the interacting system.
The effects of the interactions may be included in a local
approximation, which involves a simple integral over space, and at
each point $\mathbf{r}$, accounts for the difference in
energy--per--particle between \textit{homogeneous} noninteracting and
interacting systems of density $n(\mathbf{r})$ (for electrons, a
Hartree term is used to account for the long--range part of the
Coulomb interactions).  Moreover, for electrons at zero temperature,
this difference in energy between homogeneous systems can be described
by the well--known Wigner interpolation formula \cite{Wigner} or the
Gunnarsson--Lundqvist formula \cite{Gunnarsson}, and precise Quantum
Monte Carlo calculations are available \cite{Ceperley}.  Although the
local density approximation already achieves surprisingly high
accuracy for many electronic systems, the even higher precision
required for applications, e.g., in chemistry, motivates the ongoing
development of more sophisticated approaches.  Note that DFT is not,
in principle, a method for calculating the excitation spectra of the
systems studied, although the spectrum of the Kohn--Sham reference
system often fits the spectrum of the interacting system quite well
(DFT has even become a standard tool for evaluating band structure for
electrons in periodic crystals, although in principle it is only a
zeroth--order approximation in the context of the methods devised for
calculating such quantities, such as the GW method \cite{GWmethod}).

In response to the above--mentioned experimental developments, several
authors developed DFT methods for dilute--gas bosonic systems.  An
early attempt to develop a DFT with a high--accuracy Bogoliubov--type
treatment of the Kohn--Sham system was made in Ref.~\cite{AG95}, which
employs both the density distribution $n(\mathbf{r})$ and the
condensate amplitude $\Phi(\mathbf{r})$ as functional variables.
According to the analysis to be described here, such a high-level
treatment actually requires the use of three independent functional
variables, as discussed in Sec.~\ref{Sec:summary}.  A straightforward
application of DFT to boson systems, based on the density
$n(\mathbf{r})$ alone, was suggested by Nunes \cite{Nunes}.  For
ground states, i.e., at $T=0$, this approach results in a modified
Gross--Pitaevskii equation, containing terms which are nonlinear in
the coupling constant $g$.  It has been applied to experimentally
relevant regimes \cite{Banerjee2006}, and generalized, e.g., to
time--dependent potentials \cite{KimZubarev2003}, and to address
issues particular to strictly one--dimensional systems
\cite{Brand2004}.  The possibility of a generalization to finite
temperatures was noted by Nunes \cite{Nunes}, but it is inferior
compared to the two--fluid approach \cite{MCT97} (not to be confused
with the Landau two--fluid approach to superfluids), which was already
available at the time.  Specifically, the two--fluid approach achieves
improved accuracy by treating the condensate component, $\Phi$, and
the thermal component, $n-\left\vert \Phi\right\vert^{2}$, separately,
with the two components subject to different potentials (the interplay
between the two components can also be studied experimentally
\cite{Zawada08}).  Note, however, that Ref.~\cite{MCT97} considered
large systems with very weak inhomogeneities, for which finite--size
effects are negligible and one may assume local thermodynamic
equilibrium.  For such weakly--inhomogeneous systems, application of a
sophisticated DFT is superfluous.  Also note that the two--fluid
approach is less accurate than the field--theoretic approach in the
Popov approximation, which was applied at roughly the same time
\cite{Hutchinson97}.

Two different versions of DFT for bosons will be presented below,
based on the systematic thermodynamic approach.  One version is based
on treating the total density, $n = \langle\hat{\psi}^{\dagger}
\hat{\psi}\rangle$, and the condensate amplitude,
$\Phi=\langle\hat{\psi}\rangle$, as two density components.
Correspondingly, the Kohn--Sham reference system is a noninteracting
boson system which has the same $n(\mathbf{r})$ and $\Phi(\mathbf{r})$
distributions, and is subject not only to an effective noninteracting
potential $v_{\mathrm{ni}}(\mathbf{r})$ but also to a fictitious
potential $\eta_{\mathrm{ni}}(\mathbf{r})$ which couples directly to
the condensate amplitude $\Phi(\mathbf{r})$.  This version, which may
be called $\Phi$-DFT, reduces to the two--fluid approach of
Ref.~\cite{MCT97} in the limit of weak interactions and large systems.
It allows for inclusion of appropriate nonlinear--in--$g$ terms for
stronger interactions, as well as application to systems with
significant inhomogeneities.  The second version treats the anomalous
density $\Delta=\langle\hat{\psi}\hat{\psi}\rangle$ as a third
density, resulting in a Kohn--Sham system which is also subject to an
anomalous potential $\xi_{\mathrm{ni}}\left( \mathbf{r}\right)$, for
which a generalized Bogoliubov--type treatment is appropriate.  This
version is referred to as anomalous--DFT or A-DFT, and bears some
resemblance to the electronic DFT devised for superconducting systems
\cite{Oliveira88}.  In the limit of weak interactions, it reproduces
the Hartree--Fock--Bogoliubov model (a further approximation to which
yields the Popov model \cite{Griffin96}).

In order to apply a local density approximation for the interaction
effects, one needs results for uniform systems, as discussed above.
For dilute Bose gases, some results as a function of the density $n$
are available, at both vanishing \cite{HuangYang, Nunes} and finite
\cite{Griffin96, STB58, Popov, BlaizotRipka} temperatures.  However,
application of the advanced DFT versions discussed below requires
generalization of these results to functions not only of the density,
but also of the condensate amplitude for $\Phi$-DFT and of the
anomalous density for A-DFT. These generalized interacting systems are
analogous to spin--polarized uniform electronic systems, data for
which is in standard use within the local density approximation for
electrons.  The difference is that manipulating the condensate
amplitude $\Phi$ or the anomalous density $\Delta$ requires the use of
fictitious potentials, whereas the spin density can be modified by
subjecting the system to a physically realizable magnetic field.  As
the uniform systems which are under consideration are fictitious
anyway, and the results are obtained by theoretical methods (such as
the quantum Monte Carlo work mentioned above \cite{Ceperley}), the
realizability or not of the fields is of little importance.  Obtaining
high--accuracy results for the generalized uniform systems is beyond
the scope of the present work, and at this stage we will content
ourselves with expressions for the interaction effects which are valid
to first order in $g$ (for A-DFT, this corresponds to the
Hartree--Fock--Bogoliubov model as noted above), with one exception:
the leading--order results for A-DFT will allow us to deduce
next--to--leading--order results for $\Phi$-DFT. For dilute gases, the
first order approximation to A-DFT is wholly sufficient, except for
special cases with particularly strong interactions, which are
realizable near Feshbach resonances \cite{Inouye,Cornish}.  It is also
relevant to note that homogeneous systems with attractive interactions
(negative scattering lengths $a_{0}$) are absolutely unstable at long
wavelengths, but inhomogeneous attractive systems may have metastable
dilute--gas states, and have been studied experimentally
\cite{BoseNova}.  $\Phi$-DFT and A-DFT may be applied to such systems
with the leading--order expressions for the interactions, whereas a
higher--accuracy local--density approach is in principle unworkable,
because there can be no accurate thermodynamic results for the
relevant interacting homogeneous system (except at uninterestingly low
densities, where thermal excitations stabilize the long wavelength
perturbations).  The high--accuracy methods developed below thus have
a range of applicability which is limited primarily to systems with
repulsive interactions.

The outline of the paper is as follows.  Section~II introduces the
general formalism of DFT in the thermodynamic language.  Section~III
presents $\Phi$-DFT: it applies the principles of DFT to nonuniform
BECs, using the total density $n(\mathbf{r})$ and the condensate
amplitude $\Phi(\mathbf{r})$ as free variables.  The presentation
includes the Thomas--Fermi approximation for the Kohn--Sham system,
which is applicable when the inhomogeneities are weak, and the first
order approximation to the interaction energy.  In Sec.~IV, a more
general DFT scheme is developed, wherein apart from $n(\mathbf{r})$
and $\Phi(\mathbf{r})$, also the anomalous density,
$\Delta(\mathbf{r})$, is used as a third free variable.  In this case,
the $O(g)$ approximation leads to the Hartree--Fock--Bogoliubov
system.  Here too, the Thomas--Fermi approximation is introduced.
Section~V demonstrates one of the advantages of A-DFT, by showing how
a result for a homogeneous system, which is available with its $O(g)$
approximation can be obtained within $\Phi$-DFT only if more
complicated higher orders are included. Section~VI presents a
discussion of this comparison, concluding remarks, and suggestions for
future research.

\section{Finite--temperature Density--Functional Theory}

The purpose of the present section is to introduce the relevant
concepts of DFT. The thermodynamic approach of Ref.~\cite{Argaman} is
followed, and generalized to cases with several ``density
distributions''.  This will allow not only the total density of
particles $n(\mathbf{r})$, but also the condensate amplitude
$\Phi(\mathbf{r})$ and the anomalous density $\Delta(\mathbf{r})$, to
be used as free variables, as discussed above.

The Hamiltonian of the inhomogeneous system may be written as
$\hat{H}=\hat {H}_{\text{ni}}+\Lambda\hat{H}_{\text{int}}$ where
$\hat{H}_{\text{int}}$ includes all the interaction terms, and
$\hat{H}_{\text{ni}}$ is a noninteracting (quadratic in
field--operators) Hamiltonian, for which accurate solutions are
obtainable at an acceptable computational cost.  $\Lambda$ is a
continuous parameter specifying the strength of the interactions, with
$\Lambda=1$ for the full interacting system, and $\Lambda=0$ for the
noninteracting case.  The single--particle fields specifying the
inhomogeneity, such as the potential terms containing the external
potential $v(\mathbf{r})$ and other fields $\mathbf{B}(\mathbf{r})$
are included in $\hat{H}_{\text{ni}}$, and these couple to the
densities $n(\mathbf{r})$ and $\mathbf{m}(\mathbf{r})$ respectively.
In the interest of generality, the exact nature of
$\mathbf{B}(\mathbf{r})$ and $\mathbf{m}(\mathbf{r})$ will not be
specified yet, but as an example one may keep in mind electrons in a
magnetic field $\mathbf{B}(\mathbf{r})$, which couples to the spin
density (magnetization) $\mathbf{m}(\mathbf{r})$ \cite{AMspin}.

The focus of the present section is the thermodynamic treatment. The
statistical--physics problem of obtaining the grand potential $\Omega$ from
the Hamiltonian $\hat{H}$ will be tackled in later sections, where specifics
of the Hamiltonian for bosons will be given. In the first subsection here, the
foundation of DFT will be laid out, by explaining how the densities (rather
then the external fields) can be regarded as the free functional variables
which specify the inhomogeneous system. This is a direct generalization of
Legendre transforms, i.e., of the replacement of one free variable (e.g., the
chemical potential $\mu$) by another (e.g., the total number of particles,
$N$). The second subsection explains how, within the DFT framework, the
interacting system can be related to a specific \textquotedblleft
Kohn--Sham\textquotedblright\ noninteracting reference system, and how the
effects of interactions can be approximated, based on knowledge of homogeneous
interacting systems (the local density approximation).

\subsection{Legendre Transforms and the Hohenberg--Kohn Theorems}

Our starting point uses the grand potential, $\Omega(\left[  v(\mathbf{r}%
)-\mu,\mathbf{B}(\mathbf{r})\right]  ,T,\Lambda)$, which depends on the
temperature $T$ and the chemical potential $\mu$, as well as the specifics of
the Hamiltonian $\hat{H}$. The square brackets emphasize the functional
character of $\Omega$, i.e., the fact that its value depends on the potential
which is itself a function of position. The notation also makes explicit the
fact that the grand potential depends only on the difference $v(\mathbf{r}%
)-\mu$, not the values of $v(\mathbf{r})$ and $\mu$ separately. The
derivatives of the grand potential with respect to its functional variables
are
\begin{equation}
n(\mathbf{r})=\frac{\delta\Omega}{\delta v(\mathbf{r})}~,\quad\mathbf{m}%
(\mathbf{r})=-\frac{\delta\Omega}{\delta\mathbf{B}(\mathbf{r})}~.
\label{II n and m}%
\end{equation}
At this point, these equalities merely introduce notation for the derivatives;
the fact that $n(\mathbf{r})$ really is the density will become evident in the
statistical--physics discussion of the next section. The different signs used
here are a matter of convention, and are related to the fact that the
potential $v(\mathbf{r})$ repels the density $n(\mathbf{r})$, whereas the
\textquotedblleft magnetic field\textquotedblright\ $\mathbf{B}(\mathbf{r})$
attracts the \textquotedblleft magnetic moment density\textquotedblright%
\ $\mathbf{m}(\mathbf{r})$, as the magnetic energy density is given by
$-\mathbf{m}(\mathbf{r})\cdot\mathbf{B}(\mathbf{r})$.

We will use the fact that the grand potential $\Omega$ is concave in its
functional variables, i.e., that when it is evaluated at any two points,
$\left[  v_{1},\mathbf{B}_{1}\right]  $ and $\left[  v_{2},\mathbf{B}%
_{2}\right]  $ (at fixed $T>0$, $\Lambda$ and $\mu$), and at their midpoint
$\left[  v_{1/2},\mathbf{B}_{1/2}\right]  $ with $v_{1/2}=\frac{1}{2}%
(v_{1}+v_{2})$, $\mathbf{B}_{1/2}=\frac{1}{2}(\mathbf{B}_{1}+\mathbf{B}_{2})$,
the mean of the values obtained at the two arbitrary points is strictly
smaller than the value at the midpoint, $\frac{1}{2}(\Omega_{1}+\Omega
_{2})<\Omega_{1/2}$ (each one of these \textquotedblleft
points\textquotedblright\ is of course a set of functions of position, as
appropriate for a functional). This property, along with others which we shall
tacitly assume (e.g., differentiability of $\Omega$ for finite systems at
finite temperatures) can be proven by statistical mechanics methods
(incidentally, $\Omega$ is also concave in $\Lambda$ and $T$, but this will
not be used here). The concavity of $\Omega$ guarantees that there exists a
one--to--one relationship between the potentials and the densities. This
corresponds to the first Hohenberg--Kohn theorem \cite{HK64, Mermin65}. Thus,
a particular inhomogeneous system can be identified by its densities,
$n(\mathbf{r})$ and\ $\mathbf{m}(\mathbf{r})$, instead of specifying the
fields $v(\mathbf{r})$ and $\mathbf{B}(\mathbf{r})$.
%

\begin{figure}[th]
\centering
\includegraphics[width=0.45\textwidth]{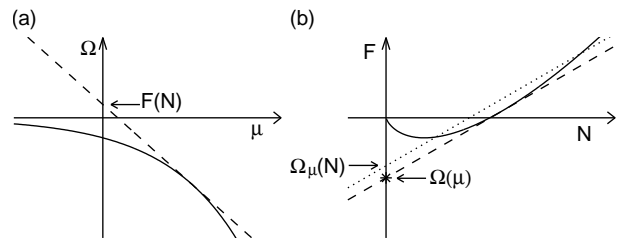}\caption{(a)
Legendre transform to obtain $F(N)$ from $\Omega( \mu)$ (see text).
(b) Legendre transform back from $F(N)$ to $\Omega(\mu)$.  The
minimization procedure for the inverse Legendre transform is
demonstrated in the second panel.}%
\label{Fig Omega(mu) and F(N)}%
\end{figure}

It is convenient to demonstrate this graphically, together with the Legendre
transforms to be introduced next, using one of the scalar variables of
$\Omega$, the chemical potential $\mu$. The corresponding partial derivative
of $\Omega$ is
\begin{equation}
N=-\frac{\partial\Omega}{\partial\mu} \label{II N}%
\end{equation}
where $N=\int d\mathbf{r\,}n(\mathbf{r})$ is the total particle number. The
one--to--one character of the relationship between $\mu$ and $N$ follows from
the monotonic dependence of the derivative of $N$ with respect to the variable
$\mu$, see Fig.~\ref{Fig Omega(mu) and F(N)}. Consider next the combination
$\Omega(\mu) +\mu N$, and maximize it over all values of $\mu$ for a given
$N$. The maximum is clearly unique, because the combination is concave in
$\mu$. By considering the derivative, one finds that at the maximum $\mu$
obeys the condition of Eq.~(\ref{II N}). This maximum value of the combination
is called the Helmholtz free energy,
\begin{equation}
F( N ) = \underset{\mu}{\;\max}\;\Omega( \mu) +\mu N = \left.  \Omega\left(
\mu\right)  +\mu N \right\vert _{\mu: N(\mu) = N} ~. \label{II F(N)}%
\end{equation}
The last equality refers to the equivalent procedure of choosing $\mu$
according to the condition of Eq.~(\ref{II N}), rather than
maximizing.  The Legendre transform from $\Omega\left( \mu\right) $ to
$F(N) $ has a simple geometric interpretation (see Fig.~\ref{Fig
Omega(mu) and F(N)}): the graph of $\Omega$ as a function of $\mu$ has
tangents of slope $-N$, and for a point $\Omega(\mu) $ on the graph,
the intercept of the tangent line with the vertical axis occurs at
$F(N) = \Omega+ \mu N$.

The function $F( N )$, describing the family of tangents (intercept as a
function of slope) to the curve $\Omega( \mu)$, contains the same information
regarding the physical system as the original function $\Omega( \mu)$, but is
in certain applications more convenient. It follows from Eq.~(\ref{II N}) that
the derivatives of $F$ are
\begin{equation}
\frac{\partial F}{\partial N}=\mu~, \quad\left(  \frac{\partial F}%
{\partial\Lambda}\right)  _{N}=\left(  \frac{\partial\Omega}{\partial\Lambda
}\right)  _{\mu}~, \label{II F derivs}%
\end{equation}
where the last equality represents a derivative with respect to a
variable not involved in the Legendre transform.  As $\mu$ increases
with $N$, the function $F( N )$ is convex (it is concave relative to
the other variables, $\Lambda$ and $T$).  An inverse Legendre
transform may thus be applied, e.g., by defining $\Omega_{\mu}(N) =
F(N) -\mu N$ and identifying the grand potential as $\Omega(\mu)
=\min_{N}\,\Omega_{\mu}(N)$ (see the right panel in the figure).  The
inverse transform differs from the original Legendre transform only in
signs.

In the case of functional variables, a geometric interpretation requires a
multitude of ``horizontal axes'' (one for each spatial point) with
high--dimensional tangent hyperplanes instead of tangent lines, but the
principle is the same. The Hohenberg--Kohn free energy of DFT is thus
introduced through a \textit{functional} Legendre transform:
\begin{multline}
F_{\text{HK}} ( \left[  n,\mathbf{m}\right]  ,T,\Lambda) = \Omega( \left[
v-\mu,\mathbf{B}\right\}  ,T,\Lambda)\label{II F HK}\\
-\int d\mathbf{r} \left\{  \left(  v-\mu\right)  n - \mathbf{B}\cdot\mathbf{m}
\right\}  ~.
\end{multline}
Here the functional variables $\left[  v-\mu,\mathbf{B}\right]  $ on the
right-hand-side (RHS) are determined by maximization, or equivalently by
requiring the physical condition of Eq.~(\ref{II n and m}) (the argument
$\mathbf{r}$ of the functions is omitted for brevity). The functional
derivatives of $F_{\text{HK}}$ are
\begin{equation}
\mu-v(\mathbf{r}) =\frac{\delta F_{\text{HK}}}{\delta n(\mathbf{r})}~,
\quad\mathbf{B}(\mathbf{r}) = \frac{\delta F_{\text{HK}}} {\delta
\mathbf{m}(\mathbf{r}) } ~. \label{II F HK derivs}%
\end{equation}
The Hohenberg--Kohn free energy, $F_{\text{HK}}( \left[  n,\mathbf{m} \right]
,T, \Lambda)$, is the generalization of the Helmholtz free energy to
inhomogeneous systems.

The inverse Legendre transform allows one to obtain the grand potential from
the Hohenberg--Kohn free energy:
\begin{multline}
\Omega\left(  \left[  v-\mu,\mathbf{B}\right]  ,T, \Lambda\right)  =\\
F_{\text{HK}}\left(  \left[  n,\mathbf{m} \right]  ,T,\Lambda\right)  +\int
d\mathbf{r}\left\{  \left(  v-\mu\right)  n-\mathbf{B}\cdot\mathbf{m}\right\}
~. \label{II Omega}%
\end{multline}
Here $n$ and $\mathbf{m}$ on the RHS are determined either by
Eq.~(\ref{II F HK derivs}) or equivalently by minimization. The second
Hohenberg--Kohn theorem corresponds to the latter statement: the RHS of
Eq.~(\ref{II Omega}), when evaluated for an interacting system ($\Lambda=1$)
at given external potentials $v = v_{\text{ext} }(\mathbf{r})$ and
$\mathbf{B}=\mathbf{B}_{\text{ext}}(\mathbf{r})$, and minimized over the
density distributions $n\left(  \mathbf{r}\right)  $ and $\mathbf{m}%
(\mathbf{r})$, gives the physical value of the grand potential $\Omega$ at the
physical density distributions. Although we will make no direct use of this
minimization principle in the following sections, relying instead on
Eq.~(\ref{II F HK derivs}), its importance in providing both a physical
picture and an avenue for developing numerical algorithms is not to be underestimated.

\subsection{The Kohn--Sham Equations}

The power of DFT stems from the feasibility of finding accurate and simple
approximations for the complicated many-body interaction effects in the free
energy $F_{\text{HK}}$. Kohn and Sham exploited the fact that the
noninteracting effects are much simpler to deal with, and nevertheless contain
the lion's share of the physics of the full system. To introduce the
Kohn--Sham scheme, the first step is to separate the Hohenberg--Kohn free
energy into two contributions:
\begin{equation}
F_{\text{HK}}(\left[  n,\mathbf{m}\right\}  ,T,\Lambda\!=\!1)=F_{\mathrm{ni}%
}(\left[  n,\mathbf{m}\right]  ,T)+F_{\text{int}}(\left[  n,\mathbf{m}\right]
,T)~, \label{II F ni F int}%
\end{equation}
where the noninteracting free energy $F_{\mathrm{ni}}$ of the Kohn--Sham
system is defined as $F_{\text{HK}}$ in the absence of interactions, i.e., at
$\Lambda=0$, and the term $F_{\text{int}}$ is defined as the difference
between the full $F_{\text{HK}}$ at $\Lambda=1$ and $F_{\mathrm{ni}}$, i.e.,
it contains all of the complicated interaction effects. In DFT for electrons,
it is standard to further separate the interaction term into a simple Hartree
long--range interaction term and an exchange--correlation term which is
usually considerably smaller, and for which approximations are sought and
employed. As will be clarified below, for neutral atoms interacting with
short--range potentials, the \textquotedblleft Hartree\textquotedblright%
\ (direct) and the \textquotedblleft exchange\textquotedblright\ contributions
are of comparable (often equal) magnitudes, and therefore we proceed with
lumping the interactions into a single term.

It follows from Eq.~(\ref{II F ni F int}) that each of the derivatives in
Eq.~(\ref{II F HK derivs}) can also be written as a sum of two terms:
\begin{align}
v_{\text{ext}}(\mathbf{r})  &  =v_{\mathrm{ni}}(\mathbf{r})-v_{\text{int}%
}(\mathbf{r})~,\label{II v eff v int}\\
\mathbf{B}_{\text{ext}}(\mathbf{r})  &  =\mathbf{B}_{\mathrm{ni}}%
(\mathbf{r})-\mathbf{B}_{\text{int}}(\mathbf{r})~,\nonumber
\end{align}
where we have used subscripts ext and ni to denote the potentials
corresponding to the $\left[  n,\mathbf{m}\right]  $ densities for $\Lambda=1$
and for $\Lambda=0$ respectively, and the interaction potentials are defined
as
\begin{equation}
v_{\text{int}}(\mathbf{r})=\frac{\delta F_{\text{int}}}{\delta n(\mathbf{r}%
)}~,\quad\mathbf{B}_{\text{int}}(\mathbf{r})=-\frac{\delta F_{\text{int}}%
}{\delta\mathbf{m}(\mathbf{r})}~, \label{II v int}%
\end{equation}
with a convention for the signs which is opposite to that of Eq.
(\ref{II F HK derivs}). The external potentials and/or fields $\left[
v_{\text{ext}},\mathbf{B}_{\text{ext}}\right]  $ are known \textit{a priori}
in standard applications, whereas the noninteracting potentials $\left[
v_{\mathrm{ni}},\mathbf{B}_{\mathrm{ni}}\right]  $ (traditionally called
effective potentials), which are required to reproduce without interactions
the same density distributions $\left[  n,\mathbf{m}\right]  $ as in the fully
interacting system, are not initially known and must be found.
Eq.~(\ref{II v eff v int}) immediately gives
\begin{align}
v_{\mathrm{ni}}(\mathbf{r})  &  =v_{\text{ext}}(\mathbf{r})+v_{\text{int}%
}(\mathbf{r})~,\label{II v eff}\\
\mathbf{B}_{\mathrm{ni}}(\mathbf{r})  &  =\mathbf{B}_{\text{ext}}%
(\mathbf{r})+\mathbf{B}_{\text{int}}(\mathbf{r})~,\nonumber
\end{align}
which gives the noninteracting or effective potentials in terms of the
externally applied fields plus a contribution due to interactions. The system
of noninteracting particles in these effective potentials serves as the
reference system for DFT calculations, and is called the Kohn--Sham system.

Eq.~(\ref{II v eff}) represents a self-consistent requirement which lies at
the heart of the Kohn--Sham scheme: given an initial guess for the density
distributions, and a practical approximation for the interaction contribution,
this relation specifies the potentials for the noninteracting reference
(Kohn--Sham) system. This reference system may then be solved using the known
tools for noninteracting particles (e.g., the single-particle Shr\"{o}dinger
equation with the Fermi--Dirac distribution for the occupations of the
electrons). The new densities may then be used as an improved guess, yielding
new values for the noninteracting potentials, in an iterative fashion. The
iterations are stopped once self--consistency has been achieved to the desired accuracy.

It remains to specify the approximation for $F_{\text{int}}$ to be used. We
will limit attention here to \textit{local density approximations} (LDAs), of
the type suggested (for electrons) by Kohn and Sham \cite{KS65}. Within this
approach, the interaction term is approximated by using the properties of
\textit{uniform} interacting systems:
\begin{equation}
F_{\text{int}}\simeq\int d\mathbf{r}\,f_{\text{int}}(n(\mathbf{r}%
),\mathbf{m}(\mathbf{r}))~. \label{II LDA}%
\end{equation}
Here $f_{\text{int}}\left(  n,\mathbf{m}\right)  $ is the contribution of
interactions to the Hohenberg--Kohn free energy of a uniform system with
densities $n$ an $\mathbf{m}$, calculated per unit volume. With this simple
expression for the interaction term, the functional derivatives defining the
contribution to the potentials, Eq.~(\ref{II v int}), can easily be taken:
\begin{align}
v_{\text{int}}(\mathbf{r})  &  =\frac{\partial f_{\text{int}}}{\partial
n}(n(\mathbf{r}),\mathbf{m}(\mathbf{r}))~,\nonumber\\
\mathbf{B}_{\text{int}}(\mathbf{r})  &  =-\frac{\partial f_{\text{int}}%
}{\partial\mathbf{m}}(n(\mathbf{r}),\mathbf{m}(\mathbf{r}))~. \label{II v LDA}%
\end{align}
Knowledge of $f_{\text{int}}(n,\mathbf{m})$ comes from outside of DFT. The
uniform system is much simpler than the non-uniform system in principle, but
evaluation of the many--body effects even in the uniform case can require
sophisticated techniques. For example, for electron systems, quantum Monte
Carlo techniques have been employed, as already noted. Once the results are
found, the function $f_{\text{int}}(n,\mathbf{m})$ can be tabulated or
otherwise efficiently represented. The results of the sophisticated
calculations for uniform systems are thus imported, using DFT, as input for
the calculations of inhomogeneous systems.

It is of interest to note that the thermodynamic derivation used here is
constructive. For example, it immediately gives the exact relation
$F_{\text{int}}=\int_{0}^{1}d\Lambda\,\left(  \partial F/\partial
\Lambda\right)  $, with the integrand $(\partial F/\partial\Lambda)$ equal to
$\partial\Omega/\partial\Lambda=\langle\hat{H}_{\text{int}}\rangle$, which in
the context of electrons has been called the adiabatic connection formula
\cite{adiabatic}, and has been derived via a much less direct route. For
weakly interacting bosons, it is appropriate to approximate the integrand here
by its noninteracting value at $\Lambda=0$. As we will see below, this yields
a particularly simple approximation for $F_{\text{int}}$, which is again
local, i.e., of the form of Eq.~(\ref{II LDA}).

\section{Density Functional Theory for Bosons --- $\Phi$-DFT}

In this section, a version of DFT adapted to bosonic systems, in which the
condensate will be treated as a separate field (the condensate field $\Phi$),
in addition to the density, will be developed. A system of identical bosonic
atoms of mass $m$ in an external potential $v_{\text{ext}}(\mathbf{r})$ can be
described, in second-quantized notation, by the Hamiltonian
\begin{equation}
\hat{H}=\int d\mathbf{r\,}\hat{\psi}^{\mathbf{\dag}}\left(  -\frac{\hbar
^{2}\nabla^{2}}{2m}+v_{\text{ext}}\right)  \hat{\psi}+\hat{H}_{\text{int}}~,
\label{III H}%
\end{equation}
where the interaction term involves a two--body interaction potential
$V(\mathbf{r-r}^{\prime})$. This potential has a hard--core repulsive form at
small interatomic separations, and a long--range attractive van-der-Waals form
outside the core. As explained in the introduction, for a dilute gas, with the
typical distance between atoms much larger than the range of the potential,
only the $s$-wave scattering contribution is significant, and the interaction
can be fully characterized by a single parameter, $a_{0}$, the $s$-wave
scattering length.

The scattering length $a_{0}$ is typically of the order of nanometers, while
the typical distance between the atoms, $n^{-1/3}$, in experiments, is of
order hundreds of nanometers. One may therefore use a Hamiltonian with a point
interaction,
\begin{equation}
\hat{H}_{\text{int}}=\frac{g}{2}\int d\mathbf{r\,}\hat{\psi}^{\mathbf{\dag}%
}\hat{\psi}^{\mathbf{\dag}}\hat{\psi}\,\hat{\psi}~, \label{III H int}%
\end{equation}
where a high--momentum cutoff $\hbar k_c$ is implied, i.e., no attempt
to describe the components of the field $\hat{\psi}$ on length--scales
as small as the range of the interaction potential is made.  Note that
physical quantities derived from this Hamiltonian depend on both the
interaction strength $g$ and the cutoff $k_c$.  For example, the
$s$-wave scattering length is related to the parameters in the
Hamiltonian by $a_{0}\simeq gm/4\pi\hbar^{2}$ only to leading order in
$g$, with corrections of order $k_c a_{0}^{2}$, which will be assumed
small.  This undesirable feature may be avoided by using a
short--range pseudo--potential \cite{HuangYang, Olshanii}, an option
which will not be made explicit here, but is necessary when large
values of $a_{0}$ are encountered (Feshbach resonances).  Note that
different forms of $\hat{H}_{\text{int}}$ are legitimate within DFT as
developed below, and are associated with different interaction
contributions $F_{\text{int}}$.  Thus, when the pseudo-potential form
of $\hat{H}_{\text{int}}$ is used, and the corresponding changes are
made in $F_{\text{int}}$, all of the DFT expressions to be derived
below will remain valid (expressions for $F_{\text{int}}$ beyond the
leading order are not included in the present work).  Furthermore, one
may include, e.g., three--body interactions, simply by modifying
$F_{\text{int}}$ appropriately.

\subsection{The Grand Potential and the Free Energy}

For a bosonic system coupled to a particle reservoir at chemical potential
$\mu$ and temperature $T$, the grand potential may be written as
\begin{equation}   \label{III Omega}
\Omega\left( \left[ v-\mu,\eta,\eta^{\ast}\right] ,T,\Lambda\right)
=-k_{\text{B}}T\,\ln\text{Tr}\exp\left(
-\frac{\hat{H}^{(\mu)}}{k_{\text{B} }T}\right) ~,
\end{equation}
where the trace is over the full many--body Hilbert space. Fictitious
potential fields, $\eta(\mathbf{r})$ and $\eta^{\ast}(\mathbf{r})$, which
break the particle--number conservation symmetry, have been included here in
order to couple to the condensate fields, $\Phi(\mathbf{r})$\ and $\Phi^{\ast
}(\mathbf{r})$ which will be introduced shortly. The grand--canonical
Hamiltonian, $\hat{H}^{(\mu)}\equiv\hat{H}-\mu\hat{N}$ with $\hat{N}$ the
number operator, is
\begin{eqnarray}  \label{III H mu}
\hat{H}^{(\mu)} &=& 
\int d\mathbf{r}\left\{  \hat{\psi}^{\mathbf{\dag}}(-\frac{\hbar^{2}\nabla
^{2}}{2m}+v-\mu)\hat{\psi}-\eta\hat{\psi}^{\mathbf{\dag}}-\eta^{\ast}\hat
{\psi}\right\}  \nonumber \\
&& + \Lambda\hat{H}_{\text{int}}~.
\end{eqnarray}
It will be convenient to treat $\eta(\mathbf{r})$ and $\eta^{\ast}
(\mathbf{r})$ (and similarly $\Phi(\mathbf{r})$ and
$\Phi^{\ast}(\mathbf{r})$, see below) as independent, and to set them
equal to the complex conjugates of each other at the end of the
calculation.  Clearly, the physical fictitious fields vanish,
$\eta_{\text{ext}}(\mathbf{r})=\eta_{\text{ext}}^{\ast
}(\mathbf{r})=0$, but the noninteracting or effective fields, $\eta
_{\mathrm{ni}}(\mathbf{r})=\eta_{\text{int}}(\mathbf{r})$, may be
significant.

The statistical--physics definition of Eq.~(\ref{III Omega}) fulfills all the
thermodynamic requirements assumed in the previous section. Specifically, it
follows directly from Eq.~(\ref{III Omega}) that $\Omega$ is concave
\cite{concave}. Its functional derivatives are as described in Eq.
(\ref{II n and m}), where we can now identify the density as
\begin{equation}  \label{III n}
n(\mathbf{r}) =\frac{\delta\Omega}{\delta v\left( \mathbf{r}\right)
}=\langle\hat{\psi}^{\mathbf{\dag}}( \mathbf{r})
\hat{\psi}(\mathbf{r}) \rangle~,
\end{equation}
and the condensate field
\begin{equation}   \label{III Phi}
\Phi(\mathbf{r}) =-\frac{\partial\Omega}{\partial\eta^{\ast}(\mathbf{r})
}=\langle\hat{\psi}( \mathbf{r} ) \rangle~,
\end{equation}
with the corresponding expression for $\Phi^{\ast}$ implied. The entropy and
interaction energy are given by
\begin{equation}
S=-\frac{\partial\Omega}{\partial T}~, \quad\mathcal{E}_{\text{int}}%
=\langle\hat{H}_{\text{int}}\rangle=\frac{\partial\Omega}{\partial\Lambda}~.
\label{III S Eint}%
\end{equation}

The principles of DFT detailed in the previous section may now be applied,
with the fictitious potential and the condensate field replacing the
\textquotedblleft magnetic\textquotedblright\ terms $\mathbf{B\left(
\mathbf{r}\right)  }$ and $\mathbf{m}(\mathbf{r})$. The Hohenberg--Kohn free
energy is thus
\begin{multline}
F_{\text{HK}}\left(  \left[  n,\Phi,\Phi^{\ast}\right]  ,T,\Lambda\right)
=\Omega\left(  \left[  v-\mu\right]  ,T,\Lambda\right) \label{III F
HK}\\
-\int d\mathbf{r\,}\left\{  \left(  v-\mu\right)  n-\eta^{\ast}\Phi-\eta
\Phi^{\ast}\right\}  ~,
\end{multline}
and its derivatives are
\begin{align}
&  \frac{\delta F_{\text{HK}}}{\delta n}=-\left(  v-\mu\right)  ~,\quad
\frac{\delta F_{\text{HK}}}{\delta\Phi}=\eta^{\ast}~,\label{III dF dn}\\
&  \frac{\partial F_{\text{HK}}}{\partial T}=-S~,\quad\frac{\partial
F_{\text{HK}}}{\partial\Lambda}=\mathcal{E}_{\text{int}}~.
\label{III F derivs}%
\end{align}

\subsection{The Kohn--Sham Equations}

We next apply the Kohn--Sham approach, based on the partition in
Eq.~(\ref{II F ni F int}) of $F_{\text{HK}}$ into a term describing a
noninteracting reference system and an interaction term.  The
noninteracting reference system, i.e., the Kohn--Sham system, is
described by the grand--canonical Hamiltonian (we drop the $\mu$
superscript to simplify notation)
\begin{equation}
\hat{H}_{\mathrm{ni}}=\int d\mathbf{r\,}\{\hat{\psi}^{\mathbf{\dag}}
(-\frac{\hbar^{2}\nabla^{2}}{2m}+v_{\mathrm{ni}}-\mu)\hat{\psi}-\eta
_{\mathrm{ni}}\hat{\psi}^{\mathbf{\dag}}-\eta_{\mathrm{ni}}^{\ast}\hat{\psi
}\}, \label{III Hni}%
\end{equation}
with the noninteracting (or effective) potentials given by
Eq.~(\ref{II v eff}).  The field operator $\hat{\psi}(\mathbf{r})$\
may be written in terms of the condensate field $\Phi(\mathbf{r})$ and
a residual operator field $\hat{\phi }(\mathbf{r})$:
\begin{equation}
\hat{\psi}(\mathbf{r})=\Phi(\mathbf{r})+\hat{\phi}(\mathbf{r})~.
\label{III phi shift}
\end{equation}
The requirement $\langle\hat{\phi}(\mathbf{r})\rangle=0$, cf.
Eq.~(\ref{III Phi}), is associated with a modified Gross--Pitaevskii
equation,
\begin{equation}
\left( -\frac{\hbar^{2}\nabla^{2}}{2m}+v_{\mathrm{ni}}-\mu\right) \Phi
=\eta_{\mathrm{ni}}~.  \label{III phi cond}
\end{equation}
This condition leads to the vanishing of the linear--in--$\hat{\phi}^{\dagger
}$ (and similarly, in $\hat{\phi}$) terms in the Hamiltonian,
\begin{multline}
\hat{H}_{\mathrm{ni}}=\hat{H}_{\text{th}}+\hat{H}_{\text{con}} =
\label{III Hni shift} \\
\int d\mathbf{r\,}\left\{  \hat{\phi}^{\mathbf{\dag}}\left(  -\frac{\hbar
^{2}\nabla^{2}}{2m}+v_{\mathrm{ni}}-\mu\right)  \hat{\phi}-\frac{1}{2}\left(
\eta_{\mathrm{ni}}\Phi^{\ast}\mathbf{+}\eta_{\mathrm{ni}}^{\ast}\Phi\right)
\right\}  ~,
\end{multline}
and hence to $\langle\hat{\phi}(\mathbf{r})\rangle=0$.  A partial
cancellation of the term involving the condensate field has occurred
here, and we have introduced notation separating ``thermal'' and
``condensate'' parts.

The Schr\"{o}dinger equation associated with the thermal part of the
Hamiltonian is
\begin{equation}
\left(  -\frac{\hbar^{2}\nabla^{2}}{2m}+v_{\mathrm{ni}}-\mu\right)
\varphi_{j}=\varepsilon_{j}\varphi_{j} \label{III Schroed}%
\end{equation}
for the single--particle wave functions $\varphi_{j}$\ and their eigenvalues
$\varepsilon_{j}$\ (i.e., the single--particle energies, measured from the
chemical potential). Using eigenstate creation and annihilation operators,%
\begin{equation}
\hat{\phi}(\mathbf{r})=\sum\limits_{j}\varphi_{j}\left(  \mathbf{r}\right)
\hat{a}_{j}~,\quad\hat{\phi}^{\mathbf{\dag}}\left(  \mathbf{r}\right)
=\sum\limits_{j}\varphi_{j}^{\ast}\left(  \mathbf{r}\right)  \hat{a}%
_{j}^{\mathbf{\dag}}~, \label{III phi a}%
\end{equation}
one may rewrite this effective noninteracting many--body Hamiltonian as%
\begin{equation}
\hat{H}_{\text{th}}=\sum\limits_{j}\varepsilon_{j}\hat{a}_{j}^{\dag}\hat
{a}_{j}~.
\end{equation}
It then becomes straightforward to evaluate the statistical--mechanical
properties of this noninteracting Kohn--Sham system. The grand potential,
Eq.~(\ref{III Omega}), also separates into two parts,
\begin{multline}
\Omega_{\mathrm{ni}}\left(  \left[  v_{\mathrm{ni}}-\mu,\eta_{\mathrm{ni}%
},\eta_{\mathrm{ni}}^{\ast}\right]  ,T\right)  =\Omega_{\text{th}}\left(
\left[  v_{\mathrm{ni}}-\mu\right]  ,T\right)  +\label{III Omega ni}\\
\Omega_{\text{con}}\left[  v_{\mathrm{ni}}-\mu,\eta_{\mathrm{ni}}%
,\eta_{\mathrm{ni}}^{\ast}\right]  ~,
\end{multline}
where
\begin{equation}
\Omega_{\text{th}}\left(  \left[  v_{\mathrm{ni}}-\mu\right]  ,T\right)
=k_{\text{B}}T\sum\limits_{j}\ln\left(  1-\exp\left(  -\frac{\varepsilon_{j}%
}{k_{\text{B}}T}\right)  \right)  ~, \label{III
Omega th}%
\end{equation}
(the requirement that the chemical potential be lower than the ground state of
the Schr\"{o}dinger equation, $\min_{j}\varepsilon_{j}>0$, is manifest here),
and
\begin{equation}
\Omega_{\text{con}}(\left[  v_{\mathrm{ni}}-\mu,\eta_{\mathrm{ni}}%
,\eta_{\mathrm{ni}}^{\ast}\right]  )=\mathbf{-\;}\frac{1}{2}\int
d\mathbf{r}\,\left(  \eta_{\mathrm{ni}}\Phi^{\ast}+\eta_{\mathrm{ni}}^{\ast
}\Phi\right)  ~. \label{III Omega cond}%
\end{equation}
The Shr\"{o}dinger equation, Eq.~(\ref{III Schroed}), determines the
eigenvalues in $\Omega_{\text{th}}$, and the Kohn--Sham form of the
Gross--Pitaevskii equation, Eq.~(\ref{III phi cond}) determines the condensate
field in $\Omega_{\text{con}}$. Note that the effective noninteracting
potential $v_{\mathrm{ni}}$ appears in both, whereas the fictitious potential
$\eta_{\mathrm{ni}}$ appears only in the latter, and that $\Omega_{\text{con}%
}$ does not depend on the temperature $T$.

Turning to the functional derivatives, one finds that the density
distribution, Eq.~(\ref{III n}), becomes%
\begin{align}
n(\mathbf{r})  &  =n_{\text{th}}(\mathbf{r})+n_{\text{con}}(\mathbf{r}%
)\nonumber\label{III n ni}\\
&  =\sum\limits_{j}\frac{\left\vert \varphi_{j}(\mathbf{r})\right\vert ^{2}%
}{\exp\left(  \varepsilon_{j}/T\right)  -1}+\Phi^{\ast}(\mathbf{r}%
)\Phi(\mathbf{r})~,
\end{align}
together with $\Phi=-\delta\Omega_{\text{con}}/\delta\eta_{\mathrm{ni}}^{\ast
}$. These relations are not only obvious from Eq.~(\ref{III phi shift}), but
can also be derived from Eq.~(\ref{III phi cond}). \ Explicitly, one takes its
variation and multiplies by $\Phi^{\ast}$ to obtain%
\begin{equation}
\Phi^{\ast}\left(  -\frac{\hbar^{2}\nabla^{2}}{2m}+v_{\mathrm{ni}}-\mu\right)
\delta\Phi+\Phi^{\ast}\Phi\delta v_{\mathrm{ni}}=\Phi^{\ast}\delta
\eta_{\mathrm{ni}}~,
\end{equation}
where the first term may be identified as $\eta_{\mathrm{ni}}^{\ast}\delta
\Phi$, and then%
\begin{equation}
\delta\Omega_{\text{con}}=\int d\mathbf{r}\,\left(  -\Phi^{\ast}\delta
\eta_{\mathrm{ni}}-\Phi\delta\eta_{\mathrm{ni}}^{\ast}+\Phi^{\ast}\Phi\delta
v_{\mathrm{ni}}\right)
\end{equation}
follows.

The Hohenberg--Kohn free energy is
\begin{equation}
F_{\mathrm{ni}}\left(  \left[  n,\Phi,\Phi^{\ast}\right]  ,T\right)
=F_{\text{th}}\left(  \left[  n-\Phi^{\ast}\Phi\right]  ,T\right)
+F_{\text{con}}\left[  \Phi,\Phi^{\ast}\right]  ~, \label{III F ni}%
\end{equation}
with
\begin{align}
F_{\text{th}}\left(  \left[  n_{\text{th}}\right]  ,T\right)   &
=k_{\text{B}}T\sum\limits_{j}\ln\left(  1-\exp\left(  -\frac{\varepsilon_{j}%
}{k_{\text{B}}T}\right)  \right) \nonumber\label{III Fth r}\\
&  -\int d\mathbf{r\,}n_{\text{th}}\left(  v_{\mathrm{ni}}-\mu\right)  ~,
\end{align}
and
\begin{equation}
F_{\text{con}}\left[  \Phi,\Phi^{\ast}\right]  =\int d\mathbf{r\,}\Phi^{\ast
}\left(  -\frac{\hbar^{2}\nabla^{2}}{2m}\right)  \Phi~. \label{III Fcond}%
\end{equation}
As is generally the case with Legendre transforms, the RHS of Eq.
(\ref{III Fth r}) is evaluated for the potential $v_{\mathrm{ni}}\left(
\mathbf{r}\right)  $ which corresponds to the given density $n_{\text{th}%
}(\mathbf{r})$, and it is difficult to make it more explicit. However, use of
Eq.~(\ref{III phi cond}) has yielded a significant simplification in the
condensate term, defined as $F_{\text{con}}=\Omega_{\text{con}}\mathbf{-\;}%
\int d\mathbf{r}\,\left(  \left(  v_{\mathrm{ni}}-\mu\right)  n_{\text{con}%
}-\eta_{\mathrm{ni}}\Phi^{\ast}-\eta_{\mathrm{ni}}^{\ast}\Phi\right)  $,
resulting in the explicit form of Eq.~(\ref{III Fcond}). The rule for Legendre
transforms of sums such as $\Omega_{\mathrm{ni}}=\Omega_{\text{th}}%
+\Omega_{\text{con}}$ is that each term can be transformed separately,
$\Omega_{\text{th}}$ into $F_{\text{th}}$ and $\Omega_{\text{con}}$ into
$F_{\text{con}}$, but the sum must be evaluated as $F_{\mathrm{ni}}\left[
n\right]  =F_{\text{th}}\left[  n_{\text{th}}\right]  +F_{\text{con}}\left[
n_{\text{con}}\right]  $ with the conditions $n=n_{\text{th}}+n_{\text{con}}$
and $\delta F_{\text{th}}/\delta n_{\text{th}}=\delta F_{\text{con}}/\delta
n_{\text{con}}$ implied. In the present case the contribution of the
condensate to the density, $n_{\text{con}}=\Phi^{\ast}\Phi$ is known in terms
of the condensate amplitude, which is itself a free variable, and no implicit
relationship remains to be evaluated. In other words, the fact that
$\delta\Omega_{\text{con}}/\delta v_{\mathrm{ni}}=\Phi^{\ast}\Phi$ is
trivially related to $\delta\Omega_{\text{con}}/\delta\eta_{\mathrm{ni}}%
^{\ast}=-\Phi$ plays a significant simplifying role, resulting in
$F_{\text{con}}$ depending only on $\Phi$ and $\Phi^{\ast}$, and
$F_{\text{th}}$ depending only on $n-\Phi^{\ast}\Phi$.

In summary, the Kohn--Sham equations for a system of bosons are Eq.
(\ref{III phi cond}) for the condensate field, Eqs.~(\ref{III Schroed}) and
(\ref{III n ni}) for the density, and $v_{\mathrm{ni}}=v_{\text{ext}%
}+v_{\text{int}}$ and $\eta_{\mathrm{ni}}=\eta_{\text{int}}$ for the effective
potentials, from Eq.~(\ref{II v eff}). For an LDA, we have
\begin{equation}
v_{\text{int}}=\partial f_{\text{int}}/\partial n~,\quad\eta_{\text{int}%
}=-\partial f_{\text{int}}/\partial\Phi^{\ast}\text{\ ,} \label{III LDA pots}%
\end{equation}
where specific expressions for the interaction energy density, $f_{\text{int}%
}(n,\Phi,\Phi^{\ast})$, will be suggested below. Once these Kohn--Sham
equations have been solved, the grand potential may be evaluated as
\begin{align}
\Omega &  =\Omega_{\mathrm{ni}}+F_{\text{int}}-\int d\mathbf{r}\left\{
\left(  v_{\mathrm{ni}}-v_{\text{ext}}\right)  n-\left(  \eta_{\mathrm{ni}%
}\Phi^{\ast}\mathbf{+}\eta_{\mathrm{ni}}^{\ast}\Phi\right)  \right\}
\nonumber\\
&  =\sum\limits_{j}k_{\text{B}}T\ln\left(  1-\exp\left(  -\frac{\varepsilon
_{j}}{k_{\text{B}}T}\right)  \right)  +\label{III
Omega KS}\\
&  \int d\mathbf{r}\left\{  \left(  f_{\text{int}}-nv_{\text{int}}\right)
+\frac{1}{2}\left(  \eta_{\text{int}}\Phi^{\ast}\mathbf{+}\eta_{\text{int}%
}^{\ast}\Phi\right)  \right\}  \text{\ }\mathbf{.}\nonumber
\end{align}
The integral here is a generalized subtraction of the double counting of the
interaction energy included in the single--particle energies, as customarily
occurs in Hartree--like schemes.

Note that for Fermions there is no condensate term, and in the
low--temperature limit, $F_{\text{ni}}$ is simply the kinetic energy $K$. The
temperature and entropy can be thought of as a correction which is necessary
at finite temperatures. For noninteracting bosons, one still has
$F_{\text{th}}=K+TS$, but both terms vanish as the temperature is lowered, and
for large condensates the zero--point kinetic energy, $F_{\text{con}}$, may
also be negligible. In such cases, one has no significant contribution to the
Hohenberg--Kohn free energy from the Kohn--Sham system, and $F\simeq
F_{\text{int}}$ in the low--temperature limit.

\subsubsection*{The Thomas--Fermi Approximation}

Many of the relevant Bose--Einstein condensate systems studied experimentally
involve a large number of bosonic atoms, in the thousands or millions, in a
smooth external potential. In such cases it is appropriate to introduce the
Thomas--Fermi approximation \cite{Dalfovo} (adapted from many electron
systems), which takes the density of single--particle states in phase space to
be $\left(  2\pi\hbar\right)  ^{-3}$, and uses the classical relationship
$\varepsilon=\frac{p^{2}}{2m}+v_{\mathrm{ni}}(\mathbf{r}) -\mu$. The local
density of states is thus approximated as
\begin{align}
d( \varepsilon,\mathbf{r})  &  =\int_{0}^{\infty}\frac{d^{3}\mathbf{p}%
}{\left(  2\pi\hbar\right)  ^{3}}\delta\left(  \varepsilon-\frac{\left\vert
\mathbf{p}\right\vert ^{2}}{2m}-v_{\mathrm{ni}}\left(  \mathbf{r}\right)
+\mu\right) \label{III TF ldos}\\
&  =\Theta\left(  \varepsilon-v_{\mathrm{ni}}(\mathbf{r}) +\mu\right)
\frac{m\sqrt{2m\left(  \varepsilon-v_{\mathrm{ni}}\left(  \mathbf{r}\right)
+\mu\right)  }}{2\pi^{2}\hbar^{3}}\text{\ } ~,\nonumber
\end{align}
and the overall density of states is given by
\begin{equation}
d\left(  \varepsilon\right)  = \sum\limits_{j}\delta( \varepsilon-
\varepsilon_{j}) = \int d\mathbf{r} \, d(\varepsilon,\mathbf{r}) ~.
\label{III TF dos}%
\end{equation}

\begin{figure}[th]
\centering
\includegraphics[width=0.45\textwidth]{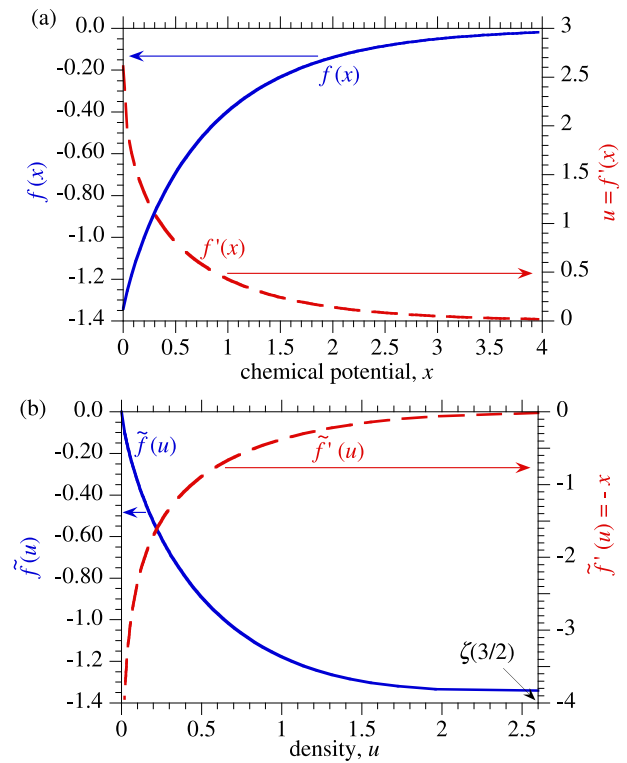}
\caption{Relationships for the thermal components according
to the Thomas--Fermi approximation: $f$ is the scaled grand potential, $x$ is
the scaled effective potential, $u$ is the scaled density, and $\tilde{f}$ is
the scaled Hohenberg--Kohn free energy. (a) The function $f(x)$ (full line),
and its derivative, $u(x)$ (dotted line). (b) The Legendre transform
$\tilde{f}(u)$ (full line), and its derivative -$x( u )$ (dotted line). Note
that the derivatives are simply inverse functions.}%
\label{Fig fff}%
\end{figure}

With this approximation, there is no need to solve the Schr\"{o}dinger
equation, Eq.~(\ref{III Schroed}), which is the step which is most significant
in terms of computational resources. Expressions such as
Eqs.~(\ref{III Omega th}) and (\ref{III n ni}) are then evaluated as simple
integrals over the corresponding density of states, Eq.~(\ref{III TF dos}) or
(\ref{III TF ldos}) respectively. For example,\ the noninteracting grand
potential, from Eqs. (\ref{III Omega th}) and (\ref{III TF dos}), becomes
\begin{equation}
\Omega_{\text{th}}=k_{\text{B}}T\lambda_{T}^{-3}\int d\mathbf{r\,}f\left(
\frac{v_{\mathrm{ni}}-\mu}{k_{\text{B}}T}\right)  ~, \label{III TF Omega ni}%
\end{equation}
where $\lambda_{T}$ is the thermal de Broglie wavelength mentioned in the
introduction, and
\begin{equation}
f(x)=\frac{4}{\sqrt{\pi}}\int_{0}^{\infty}q^{2}dq\,\ln\left(  1-\text{e}%
^{-q^{2}-x}\right)  =-\sum_{l=1}^{\infty}\frac{\text{e}^{-lx}}{l^{5/2}%
}\text{\ ,} \label{III f(x)}%
\end{equation}
with $x=(v_{\mathrm{ni}}-\mu) /k_{\text{B}}T$, and $q$ a scaled
momentum variable.  The function $f(x)$ is known as the {\em
polylogarithm} or {\em de Jonquiere's function}, and is plotted in
Fig.~\ref{Fig fff}.  It varies from $-\zeta(5/2)$ to $0$ as $x$ is
varied from $0$ to $\infty$ (the Riemann zeta--function evaluates to
$\zeta(5/2)=1.341\ldots$).  Its derivative (also plotted), which
varies from $\zeta(3/2)=2.612\ldots$ to 0, determines the density,
\begin{equation}
n_{\text{th}}(\mathbf{r})=\lambda_{T}^{-3}f^{\prime}\left(  \frac
{v_{\mathrm{ni}}-\mu}{k_{\text{B}}T}\right)  ~. \label{III TF n f'}%
\end{equation}
Its Legendre transform (cf. the figure again) is
\begin{equation}
\tilde{f}(u)=\max_{x}\{f(x)-ux\}~, \label{III f tilde}%
\end{equation}
with $u=\lambda_{T}^{3}n_{\text{th}}$, the dimensionless density, and this
determines the noninteracting Hohenberg--Kohn free energy as
\begin{equation}
F_{\text{th}}(\left[  n_{\text{th}}\right]  ,T)=k_{\text{B}}T\lambda_{T}%
^{-3}\int d\mathbf{r\,}\tilde{f}(\lambda_{T}^{3}n_{\text{th}})~.
\label{III Fth TF}%
\end{equation}

The Thomas--Fermi approximation is appropriate for systems with gradual
inhomogeneities. It may be applied to the condensate component as well, by
simply dropping the gradient term in Eq.~(\ref{III phi cond}), which amounts
to neglecting the zero--point energy of the condensate, $F_{\text{con}}
\simeq0$ (this corresponds to $\Omega_{\text{con}} \simeq\int d\mathbf{r}
(v_{\mathrm{ni}}-\mu) \Phi^{\ast}\Phi$ in the above notation). As mentioned in
the introduction, when all the finite--size effects due to the inhomogeneities
in the system are indeed negligible, it is appropriate to use a
local--equilibrium approach, with the density $n(\mathbf{r})$ at each position
taken as that which for an infinite system would correspond to the given local
value of the chemical potential, $\mu-v_{\text{ext}}(\mathbf{r})$.
Applications of DFT to such situations approach the local--equilibrium
results. For $\Phi$-DFT, the condensate amplitude $\Phi(\mathbf{r})$ relaxes
to the value corresponding to an infinite system of density $n(\mathbf{r})$,
and there is thus no point in including it as a separate functional variable.

\subsection{Interaction Effects}

In order to complete the DFT description, an approximate description of the
interaction term, $F_{\text{int}}(\left[  n,\Phi,\Phi^{\ast}\right]  ,T)$ must
be specified. As noted at the end of Sec. II, the simplest approximation is
obtained by equating the integrand in the adiabatic connection formula with
its value for noninteracting bosons:
\begin{equation}
F_{\text{int}}(\left[  n,\Phi,\Phi^{\ast}\right]  ,T)\simeq\int d\mathbf{r\,}%
\frac{g}{2}\left(  2n^{2}-\left(  \Phi^{\ast}\Phi\right)  ^{2}\right)  ~,
\label{III Fint 1st O}%
\end{equation}
or $f_{\text{int}}=\left(  g/2\right)  \left(  2n^{2}-\left(  \Phi^{\ast}%
\Phi\right)  ^{2}\right)  $ in the notation of Eq. (\ref{II LDA}). The factor
of 2 in the brackets comes from counting both the direct and the exchange
contributions, and the subtraction comes from the fact that exchange is\ not
relevant to the condensate's interaction with itself. From
Eq.~(\ref{III LDA pots}), this leads to
\begin{equation}
v_{\text{int}}=2gn~,\quad\eta_{\text{int}}=g\left(  \Phi^{\ast}\Phi\right)
\Phi\text{\ ,} \label{III 1st O pots}%
\end{equation}
or to
\begin{equation}
v_{\mathrm{ni}}=v_{\text{ext}}+2gn~,\quad\eta_{\mathrm{ni}}=g\left(
\Phi^{\ast}\Phi\right)  \Phi\text{\ .} \label{III 1st O eff pots}%
\end{equation}
Using the latter in Eq, (\ref{III phi cond}) gives the Gross--Pitaevskii
equation,
\begin{equation}
\left(  -\frac{\hbar^{2}\nabla^{2}}{2m}+v_{\text{ext}}-\mu+2gn_{\text{th}%
}+g\Phi^{\ast}\Phi\right)  \Phi=0~. \label{III GP}%
\end{equation}
As mentioned in the introduction, the interaction terms here differ from the
simple $2gn$ appearing in the effective potential, due to the absence of an
exchange contribution to the condensate--condensate interactions. The
first--order approximation of Eq.~(\ref{III Fint 1st O}), with $f_{\text{int}%
}$ quadratic in $n$, $\Phi$ and $\Phi^{\ast}$, leads to simplification of Eq.
(\ref{III Omega KS}) for the grand potential, resulting in $\Omega
=\Omega_{\text{th}}-\int f_{\text{int}}d\mathbf{r}$, where the subtraction of
the double counting of the interaction energy is explicit. This simplification
is not as dramatic as it may seem, as the subtraction can only be evaluated
after the Kohn--Sham system of equations has been solved (either with or
without the Thomas--Fermi approximation for the thermal cloud and for the 
condensate).

Note that Eq.~(\ref{III GP}) is identical to the G-P equation derived
from the field-theory approach in the Popov approximation
\cite{Griffin96}.

The present level of description, with the Thomas--Fermi approximation for the
density of the thermal component, Eq.~(\ref{III TF n f'}), reproduces the
two--fluid\ description of finite--temperature Bose--Einstein condensates
\cite{MCT97} mentioned in the introduction. The present $\Phi$-DFT provides a
route for improvements in this description, based on improved evaluations of
$f_{\text{int}}(n,\Phi,\Phi^{\ast},T)$. In fact, we will see in Sec. V that
such improvements can be appreciable even when the interactions are not
particularly strong. Furthermore, $\Phi$-DFT allows treatment of systems with
significant inhomogeneities, which are not describable by the simple
two--fluid equations.

Before closing this section, it is appropriate to state explicitly the
differences in treatment which obtain for a DFT of bosons with only a single
density. At zero temperature, one has $n_{\text{th}}=0$ or $n=\Phi^{\ast}\Phi
$, a single--density treatment would have $f_{\text{int}}(n) = gn^{2}/2$, and
the Gross--Pitaevaskii equation [Eq.~(\ref{III GP}) without $n_{\text{th}}$]
arises as the ground--state solution of the Schr\"{o}dinger equation, and need
not be derived by shifting the quantum operator as in Eq.~(\ref{III phi cond}%
). It is thus seen that in this limit the present treatment does not differ
significantly from the single--density DFT treatment suggested by Nunes
\cite{Nunes}. Substantial differences do arise at finite temperatures, where a
single--density treatment with a first--order local approximation would have
$f_{\text{int}}= (g/2) \left(  2n^{2}-\left(  n - \zeta(3/2)\lambda_{T}%
^{-3}\right)  ^{2}\right)  $, and the corresponding effective potential,
$v_{\text{eff} } = v_{\text{ext}}+g\left(  n+n_{\text{th}}\right)  $ with
$n_{\text{th}} = \zeta(3/2) \lambda_{T}^{-3}$, would still give rise to an
equation for the ground state which is essentially the correct
Gross--Pitaevskii equation, but the excited states would ``feel the wrong
potential''. In the limit of weak inhomogeneities, the situation can be
remedied. The Thomas--Fermi approximation holds, with $v_{\mathrm{ni}}-\mu=0$
at points with a condensate, i.e., with $n>\zeta(3/2) \lambda_{T}^{-3}$. The
corresponding free energy function is $\tilde{f}\left(  u\right)  $ with
$u=\lambda_{T}^{3}n$, and is to be continued to large densities, $u >
\zeta(3/2)$. According to the rules for Legendre transforms,
Eq.~(\ref{III f tilde}) it is simply linear in this regime. The solution of
the Kohn--Sham system of equations for points with $v_{\mathrm{ni}}%
(\mathbf{r}) =\mu$ would then seem to be ambiguous, as there is a range of
densities for a single value of the effective potential, but the condition
$v_{\text{ext}}(\mathbf{r}) +v_{\text{int}}( n(\mathbf{r})) =\mu$ may be used
to determine the density $n(\mathbf{r})$ instead. If the interaction energy
$f_{\text{int} }(n)$ correctly accounts for the difference between the
reference system and the interacting system, then the correct results for the
free energy and the density distribution are guaranteed to obtain. It is only
in the presence of significant inhomogeneities that the weakness of this
approach (i.e., the effect of its having essentially the same potential in the
Schr\"{o}dinger and the Gross--Pitaevskii equation) will show up.

\section{Density Functional Theory for Bosons with Anomalous Terms --- A-DFT}

In this section, the thermodynamic approach will be used to develop another
version of DFT for bosonic systems, which results from adding a term of the
form $-\int d\mathbf{r}\left[  \xi\hat{\psi}^{\dagger} \hat{\psi}^{\dagger} +
\xi^{\ast}\hat{\psi}\hat{\psi}\right]  $ to the Hamiltonian. Here
$\xi(\mathbf{r})$ is a second fictitious potential --- an anomalous potential
--- which is to be set equal to zero in the fully interacting system,
$\xi_{\text{ext}}=0$. The fact that it does not vanish in the Kohn--Sham
reference system, $\xi_{\mathrm{ni}} \neq0$, will result in a level of
treatment generalizing that of Bogoliubov. In order to assist the reader, the
partitioning into subsections here is precisely parallel to that of the above
section presenting $\Phi$-DFT.

\subsection{The grand potential and the free energy}

The thermodynamic treatment of the enlarged Hamiltonian follows the same steps
as above, with Eq.~(\ref{III Omega}) defining the grand potential, which
acquires a $\left[  \xi,\xi^{\ast} \right]  $ dependence. The corresponding
derivative is
\begin{equation}
\label{IV Delta}\Delta(\mathbf{r})=-{\frac{\delta\Omega}{\delta\xi^{\ast
}(\mathbf{r})} }=\langle\hat{\psi}(\mathbf{r})\hat{\psi}(\mathbf{r}) \rangle~,
\end{equation}
where $\xi(\mathbf{r})$ and $\xi^{\ast}(\mathbf{r})$ as well as $\eta
(\mathbf{r})$ and $\eta^{\ast}(\mathbf{r})$ play the role of $\mathbf{B}
(\mathbf{r})$. The Legendre transform leading to the Hohenberg--Kohn free
energy is
\begin{multline}
F_{\text{HK}}([n,\Phi,\Phi^{\ast},\Delta,\Delta^{\ast}],T,\Lambda
)\;=\;\label{IV F HK}\\
\Omega([v-\mu,\eta,\eta^{\ast},\xi,\xi^{\ast}],T,\Lambda)\\
\qquad-\int d\mathbf{r}\,\left\{  (v-\mu)n - \eta\Phi^{\ast} - \eta^{\ast}
\Phi- \xi\Delta^{\ast}-\xi^{\ast}\Delta\right\}  ~,
\end{multline}
and we have the additional relation
\begin{equation}
{\frac{\delta F_{\text{HK}}}{\delta\Delta}}=\xi^{\ast}\;. \label{IV F HK der}%
\end{equation}

\subsection{The Bogoliubov Kohn--Sham System}

The Kohn--Sham reference system is described here by the noninteracting
Hamiltonian
\begin{multline}
\hat{H}_{\mathrm{ni}}=\int d\mathbf{r}\,\hat{\psi}^{\dagger}\left(
\!-\frac{\hbar^{2}\nabla^{2}}{2m}+v_{\mathrm{ni}}-\mu\!\right)  \hat{\psi
}-\label{IV Hni}\\
\qquad\int d\mathbf{r}\left\{  \eta_{\mathrm{ni}}\hat{\psi}^{\dagger}%
+\eta_{\mathrm{ni}}^{\ast}\hat{\psi}+\xi_{\mathrm{ni}}\hat{\psi}^{\dagger}%
\hat{\psi}^{\dagger}+\xi_{\mathrm{ni}}^{\ast}\hat{\psi}\hat{\psi}\right\}  ~,
\end{multline}
where the noninteracting effective potential $v_{\mathrm{ni}}$ and auxiliary
fields $\eta_{\mathrm{ni}}$, $\xi_{\mathrm{ni}}$, are again to be defined by
Eq.~(\ref{II v eff}) and determined by the interactions. Shifting the field
operator by a scalar as in Eq.~(\ref{III phi shift}), $\hat{\psi}%
(\mathbf{r})=\Phi(\mathbf{r})+\hat{\phi}(\mathbf{r})$, and requiring all terms
linear in the operators $\hat{\phi}^{\dagger}$ (and $\hat{\phi}$) to cancel
from the Hamiltonian, yields in this case
\begin{equation}
\left(  -\frac{\hbar^{2}\nabla^{2}}{2m}+v_{\mathrm{ni}}-\mu\right)  \Phi
-\eta_{\mathrm{ni}}-2\xi_{\mathrm{ni}}\Phi^{\ast}=0~. \label{IV phi cond}%
\end{equation}
The Kohn--Sham Hamiltonian becomes
\begin{equation}
\hat{H}_{\mathrm{ni}}=\hat{H}_{\text{nc}}+\hat{H}_{\text{con}}~,
\label{IV H=H+H}%
\end{equation}
with
\begin{equation}
\hat{H}_{\text{nc}}=\int d\mathbf{r}\,\{\hat{\phi}^{\dagger}(-\frac{\hbar
^{2}\nabla^{2}}{2m}+v_{\mathrm{ni}}-\mu)\hat{\phi}-\xi_{\mathrm{ni}}\hat{\phi
}^{\dagger}\hat{\phi}^{\dagger}-\xi_{\mathrm{ni}}^{\ast}\hat{\phi}\hat{\phi
}\}~, \label{IV Hnc}%
\end{equation}
and $\hat{H}_{\text{con}}=-{\frac{1}{2}}\int d\mathbf{r}(\eta_{\text{eff}}%
\Phi^{\ast}+\eta_{\mathrm{ni}}^{\ast}\Phi)$ as before, Eq.
(\ref{III Hni shift}). The subscript nc represents the non-condensed part of
the boson system, which persists to zero temperature, and should thus not be
referred to as a thermal component. Note the complete cancellation of terms of
type $\int d\mathbf{r}\>\xi_{\mathrm{ni}}^{\ast}\Phi^{2}$ in the effective
Hamiltonian -- the contributions from $\Phi^{\ast}\left(  \dots\right)  \Phi$
and $\xi_{\mathrm{ni}}^{\ast}\Phi^{2}$ are equal and opposite, due to
Eq.~(\ref{IV phi cond}). In contrast, the terms of type $\int d\mathbf{r}%
\>\eta_{\mathrm{ni}}\Phi^{\ast}$ only partially cancel, leading to the
${\frac{1}{2}}$ prefactor in $\hat{H}_{\text{con}}$.

The Hamiltonian $\hat{H}_{\text{nc}}$ is quadratic in the field operators, but
does not conserve particle number. This form of Hamiltonian is diagonalized by
the Bogoliubov transformation \cite{Bog}. Following Fetter's notation
\cite{Fetter98}, the field operators may be written as
\begin{equation}
\hat{\phi}(\mathbf{r})=\sum_{j}\!^{\prime}u_{j}(\mathbf{r}) \hat{\gamma}%
_{j}-v_{j}^{\ast}(\mathbf{r})\hat{\gamma}_{j}^{\dagger} \label{IV gammas}%
\end{equation}
and its Hermitian conjugate, where the primed sum runs only over positive
energy solutions, $\mathcal{E}_{j}>0$. The $\gamma_{j}^{\dagger}$ and
$\gamma_{j}$ are bosonic creation and annihilation operators for the
Bogoliubov excitations of the system. The generalized Schr\"{o}dinger equation
for the $(u_{j},v_{j})$ wave functions is given by
\begin{align}
\left[  -\frac{\hbar^{2}\nabla^{2}}{2m}+v_{\mathrm{ni}}-\mu\right]  u_{j}
-2\xi_{\mathrm{ni}}v_{j}  &  =\mathcal{E}_{j}u_{j}\label{IV BdG}\\
\left[  -\frac{\hbar^{2}\nabla^{2}}{2m}+v_{\mathrm{ni}}-\mu\right]  v_{j}
-2\xi_{\mathrm{ni}}^{\ast}u_{j}  &  =-\mathcal{E}_{j}v_{j} ~.\nonumber
\end{align}
This two--component system of equations, known as the Bogoliubov--de Gennes
equation \cite{Fetter98}, is of the type ${\mathcal{H}}\binom{u}{v} =
\mathcal{E}\sigma_{z}\binom{u}{v}$ where ${\mathcal{H}}$ is a Hermitian matrix
differential operator, with inner product $\int d\mathbf{r} \, (u_{i}^{\ast}
\, v_{i}^{\ast}) \sigma_{z} \binom{u_{j}}{v_{j}} = \int d\mathbf{r} \,
(u_{i}^{\ast}u_{j}-v_{i}^{\ast}v_{j})$ involving the Pauli matrix $\sigma_{z}
= \left(  \! \! {
\begin{array}
[c]{cc}
1 & 0\\
0 & { - 1}
\end{array}
} \! \! \right)  $. The normalization is $\int d\mathbf{r} \, (|u_{j}%
|^{2}-|v_{j}|^{2} )=1$, and the orthogonality conditions are $\int
d\mathbf{r}\>(u_{i}^{\ast}u_{j}-v_{i}^{\ast}v_{j})=0$ for $i\neq j$ and $\int
d\mathbf{r}\>(u_{i} ^{\ast}v_{j}^{\ast}-v_{i}^{\ast}u_{j}^{\ast})=0$ for all
$i$ and $j$ with $\mathcal{E}_{i},\mathcal{E}_{j}>0$ \cite{neg_energies}.

Substituting Eq.~(\ref{IV gammas}) into Eq.~(\ref{IV Hnc}), the Hamiltonian
becomes
\begin{equation}
\hat{H}_{\text{nc}}=\sum_{j}\mathcal{E}_{j}\left(  \hat{\gamma}_{j}^{\dagger
}\hat{\gamma}_{j}-\int d\mathbf{r}|v_{j}|^{2}\right)  ~, \label{IV Hnc s}%
\end{equation}
which is now in the form of a simple harmonic oscillator for each excitation
mode $j$. With Eq.~(\ref{IV Hnc s}), expectation values of different
quantities at a temperature $T$ can be evaluated, using either
Eq.~(\ref{IV gammas}) with $\langle\hat{\gamma}_{j}\rangle=\langle\hat{\gamma
}_{j}^{\dagger}\rangle=0$ and $\langle\hat{\gamma}_{i}^{\dagger}\hat{\gamma
}_{j}\rangle=\delta_{ij}(\exp(\mathcal{E}_{j}/T)-1)^{-1}$, etc., or by
explicitly calculating the partition function and the grand potential, and
taking its derivatives. One finds that $\langle\hat{\psi}(\mathbf{r}%
)\rangle=\Phi(\mathbf{r})$, the density is $n(\mathbf{r})=n_{\text{nc}%
}(\mathbf{r})+\Phi^{\ast}\Phi$, with
\begin{equation}
n_{\text{nc}}(\mathbf{r})=\langle\hat{\phi}^{\dagger}\hat{\phi}\rangle
=\sum_{j}\left(  {\frac{|u_{j}|^{2}+|v_{j}|^{2}}{\exp(\mathcal{E}%
_{j}/k_{\text{B}}T)-1}}+|v_{j}|^{2}\right)  ~, \label{IV n nc}%
\end{equation}
and the anomalous density is $\Delta(\mathbf{r})=\Delta_{\text{nc}}%
(\mathbf{r})+\Phi^{2}$, with
\begin{equation}
\Delta_{\text{nc}}(\mathbf{r})=\langle\hat{\phi}\hat{\phi}\rangle=-\sum
_{j}u_{j}v_{j}^{\ast}\left(  {\frac{2}{\exp(\mathcal{E}_{j}/k_{\text{B}}T)-1}%
}+1\right)  ~. \label{IV Delta nc}%
\end{equation}

The grand potential of the Kohn--Sham system is $\Omega_{\mathrm{ni}}%
=\Omega_{\text{nc}}+\Omega_{\text{con}}$, with
\begin{multline}
\Omega_{\text{nc}}(\left[  v_{\mathrm{ni}}-\mu,\xi_{\mathrm{ni}}%
,\xi_{\mathrm{ni}}^{\ast}\right]  ,T)=\\
\sum_{j}\left(  k_{\text{B}}T\ln\left[  1-\exp\left(  -\frac{\mathcal{E}_{j}%
}{k_{\text{B}}T}\right)  \right]  -\mathcal{E}_{j}\int d\mathbf{r}%
\>|v_{j}|^{2}\right)  ~,\qquad\label{IV Omega nc}%
\end{multline}
and $\Omega_{\text{con}}[v_{\mathrm{ni}}-\mu,\eta_{\mathrm{ni}},\eta
_{\mathrm{ni}}^{\ast},\xi_{\mathrm{ni}},\xi_{\mathrm{ni}}^{\ast}]=-{\frac
{1}{2}}\int d\mathbf{r}(\eta_{\mathrm{ni}}\Phi^{\ast}+\eta_{\mathrm{ni}}%
^{\ast}\Phi)$ as in Eq.~(\ref{III Omega cond}). The Hohenberg--Kohn free
energy, according to Eqs.~(\ref{IV F HK}) and (\ref{IV phi cond}), is
$F_{\mathrm{ni}}=F_{\text{nc}}+F_{\text{con}}$, with
\begin{multline}
F_{\text{nc}}(\left[  n-\Phi^{\ast}\Phi,\Delta-\Phi^{2},\Delta^{\ast}%
-\Phi^{\ast2}\right]  ,T)=\\
\sum_{j}\left(  k_{\text{B}}T\ln(1-\exp(-\mathcal{E}_{j}/k_{\text{B}%
}T))-\mathcal{E}_{j}\int d\mathbf{r}\>|v_{j}|^{2}\right) \label{IV Fnc}\\
-\int d\mathbf{r}\left\{  n_{\text{nc}}(v_{\mathrm{ni}}-\mu)-\Delta
_{\text{nc}}\xi_{\mathrm{ni}}^{\ast}-\Delta_{\text{nc}}^{\ast}\xi
_{\mathrm{ni}}\right\}  ~,
\end{multline}
and $F_{\text{con}}[\Phi,\Phi^{\ast}]=\int
d\mathbf{r}\,\Phi^{\ast}\left( -\frac{\hbar^{2}\nabla^{2}}{2m}\right)
\Phi$ as in Eq.~(\ref{III Fcond}).  The RHS of Eq.~(\ref{IV Fnc}) is
evaluated, as before, with the potentials $v_{\mathrm{ni}}$,
$\xi_{\mathrm{ni}}$ and $\xi_{\text{eff}}^{\ast}$ which reproduce the
non-condensate parts of the density and the anomalous density, through
Eqs.~(\ref{IV BdG}), (\ref{IV n nc}) and (\ref{IV Delta nc}).  Notice
that the non-condensate contribution to each of the thermodynamic
quantities $\Omega$, $n$, $\Delta$ and $F$ can be further divided into
a temperature--dependent thermal part and an athermal part, e.g.,
$\Omega _{\text{nc}}=\Omega_{\mathrm{th}}+\Omega_{\text{ath}}$ in
Eq.~(\ref{IV Omega nc}).  The temperature dependence yields an
exponential convergence of the thermal parts, and only the athermal
parts depend on the cutoff $k_{c}$ substantially (the above-mentioned
rule for the evaluation of a Legendre transforms of a sum of two
functions applies for $F_{\text{nc}}=F_{\text{th}}+F_{\text{ath}}$,
with requirements such as $\delta F_{\mathrm{th}}/\delta
n_{\mathrm{th}}=\delta F_{\text{ath}}/\delta n_{\text{ath}}$ and
$n_{\mathrm{th}}+n_{\text{ath}}=n_{\text{nc}}$ implied).

In summary, the Kohn--Sham equations of A-DFT are Eq.~(\ref{IV phi cond}) for
the condensate amplitude and Eq.~(\ref{IV BdG}) for the non-condensate
eigenstates and eigenvalues, together with the corresponding expressions for
the density and the anomalous density, Eqs.~(\ref{IV n nc}) and
(\ref{IV Delta nc}), and together with the self--consistent determination of
the effective potentials through Eqs.~(\ref{II v eff}) and (\ref{II v int}).
The interaction contribution to these potentials will be made explicit below.
Once this system of equations has been solved, one may use the results to
obtain the grand potential for the interacting system, which evaluates to
\begin{align}
\label{IV Omega tot}\Omega &  = \Omega_{\text{nc}} + \int d\mathbf{r} \left\{
\mathbf{\,}f_{\text{int}}-nv_{\text{int}}+\Delta\xi_{\text{int} }^{\ast
}+\Delta^{\ast}\xi_{\text{int}}+ \right. \nonumber\\
&  \left.  \left(  \eta_{\text{int} }\Phi^{\ast}+\eta_{\text{int}}^{\ast}%
\Phi\right)  /2 \right\}  ~,
\end{align}
in full analogy with Eq.~(\ref{III Omega KS}) of $\Phi$-DFT.

\subsubsection*{Thomas--Fermi Approximation for A-DFT}

For applications involving a large number of bosons, a Thomas--Fermi type of
approximation can be formulated also in the presence of the anomalous
potential. It is convenient to refer to a momentum variable $\mathbf{p} =
\hbar\mathbf{k}$, with the density of states in the single-particle phase
space taken as $(2\pi\hbar)^{-3}$, as above. The corresponding ``local'' wave
function, $\left(  u_{\mathbf{k}},v_{\mathbf{k}} \right)  \exp(i\mathbf{k}%
\cdot\mathbf{r})$, consists of plane waves with a ``bare'' energy of
$\varepsilon_{\mathbf{k}}(\mathbf{r})=( \hbar^{2} k^{2}/2m) +v_{\mathrm{ni}%
}(\mathbf{r})-\mu$ (including the position dependence due to the effective
potential). Eq.~(\ref{IV BdG}) then takes the form,
\begin{equation}
\left(  \! \!
\begin{array}
[c]{cc}%
\varepsilon_{\mathbf{k}} & -2\xi_{\mathrm{ni}}\\
-2\xi_{\mathrm{ni}}^{\ast} & \varepsilon_{\mathbf{k}}%
\end{array}
\! \! \right)  \left(  \! \!
\begin{array}
[c]{c}%
u_{\mathbf{k}}\\
v_{\mathbf{k}}%
\end{array}
\! \! \right)  =\mathcal{E}_{\mathbf{k}}\sigma_{z}\,\left(  \! \!
\begin{array}
[c]{c}%
u_{\mathbf{k}}\\
v_{\mathbf{k}}%
\end{array}
\! \! \right)  ~, \label{IV TF bog}%
\end{equation}
with the appropriate continuum normalization $|u_{\mathbf{k}}|^{2}%
-|v_{\mathbf{k}}|^{2}=1$. Solving this eigensystem of equations gives
\begin{align}
u_{\mathbf{k}}  &  =\cosh\theta_{\mathbf{k}}\,,\nonumber\\
v_{\mathbf{k}}  &  =\left(  \xi_{\mathrm{ni}}^{\ast}/|\xi_{\mathrm{ni}%
}|\right)  \sinh\theta_{\mathbf{k}}\>\,,\label{IV TF Euv}\\
\mathcal{E}_{\mathbf{k}}  &  =\varepsilon_{\mathbf{k}}/\cosh2\theta
_{\mathbf{k}}=\sqrt{\varepsilon_{\mathbf{k}}^{2}-4|\xi_{\mathrm{ni}}|^{2}%
}\,,\nonumber
\end{align}
where $\tanh2\theta_{\mathbf{k}}=2|\xi_{\mathrm{ni}}|/\varepsilon_{\mathbf{k}%
}$.

When the non-condensate parts of the thermodynamic quantities are expressed in
terms of the solutions of the Bogoliubov--de Gennes equation, they naturally
have thermal and athermal parts, as in Eq.~(\ref{IV Omega nc}), and in Eqs.
(\ref{IV n nc}) and (\ref{IV Delta nc}). It will be convenient here to
introduce dimensionless functions $f_{\mathrm{th}}$ and $f_{\text{ath}}$\ for the
corresponding contributions to the grand potential, within the Thomas--Fermi
approximation:
\begin{align}
\label{IV fT fA}\Omega_{\text{nc}}  &  \simeq k_{\text{B}}T\lambda_{T}%
^{-3}\int d\mathbf{r} \, f_{\text{th}} \left(  \frac{v_{\mathrm{ni}}-\mu}
{k_{\text{B}}T}, \frac{\left\vert \xi_{\mathrm{ni}}\right\vert }{k_{\text{B}%
}T} \right)  +\nonumber\\
&  \mathcal{E}_{c}k_{c}^{3}\int d\mathbf{r} \, f_{\text{ath}} \left(
\frac{v_{\mathrm{ni}}-\mu} {\mathcal{E}_{c}} , \frac{\left\vert \xi
_{\mathrm{ni}}\right\vert }{\mathcal{E}_{c}} \right)  ~,
\end{align}
where $\mathcal{E}_{c}=\hbar^{2}k_{c}^{2}/2m$\ is the cutoff energy,
$\left\vert \xi_{\mathrm{ni}}\right\vert $ is used as shorthand for $\sqrt
{\xi_{\text{eff}}^{\ast}\xi_{\mathrm{ni}}}$, and the functions are defined as
\begin{equation}
f_{\text{th}}(x,y)={\frac{4}{\sqrt{\pi}}}\int q^{2}dq\,\ln\left(
1-\exp(-\sqrt{(q^{2}+x)^{2}-4y^{2}})\right)  \qquad\label{IV fT}%
\end{equation}
and
\begin{equation}
f_{\text{ath}}(\bar{x},\bar{y})={\frac{2}{\sqrt{\pi}}}\int\limits_{0}^{1}%
\bar{q}^{2}d\bar{q}\left(  \sqrt{(\bar{q}^{2}+\bar{x})^{2}-4\bar{y}^{2}}%
-\bar{q}^{2}-\bar{x}\right)  \label{IV fA}%
\end{equation}
[the notation $x$, $y$, $\bar{x}$ and $\bar{y}$ will be used below as
shorthand for the corresponding combinations in Eq.~(\ref{IV fT fA})]. The
grand potential of the Kohn--Sham system is thus the sum of three terms, a
condensate part, a thermal part, and an athermal part. As the integral of the
athermal part is divergent, we have used the cutoff scale to express it in
dimensionless terms (the cutoff dependence of the thermal part is
exponentially small, and has been ignored). The cutoff energy is large, and it
will be appropriate to expand $f_{\text{ath}}$ for small values of its
variables --- see the appendix. In contrast, the temperature may be small, and
so the whole range of $f_{\text{th}}$ will be relevant, except in specific
cases such as at $T=0$.

The square root in the integrand of Eqs.~(\ref{IV fT}) and (\ref{IV fA}) is
the normalized energy $\mathcal{E}_{\mathbf{k}}$, and for $x=2y$ or
$v_{\mathrm{ni}}-\mu=2\left\vert \xi_{\mathrm{ni}}\right\vert $ it has a
linear dependence at small wavenumbers. This represents the phonon branch of
the excitation spectrum of the superfluid. The present description allows also
for situations with $x>2y$, which possess a gap in the spectrum at
$\mathbf{k}=0$. As will be discussed further below, this gap is not physical.

The Thomas--Fermi results for the density and the anomalous density are:%
\begin{align}
n_{\text{nc}}(\mathbf{r}) &  \simeq\lambda_{T}^{-3}u_{\text{th}}%
(x,y)+k_{c}^{3}u_{\text{ath}}(\bar{x},\bar{y})~,\nonumber\\
\Delta_{\text{nc}}(\mathbf{r}) &  \simeq\frac{1}{2}\sqrt{\frac{\xi
_{\mathrm{ni}}}{\xi_{\mathrm{ni}}^{\ast}}}\left(  \lambda_{T}^{-3}%
w_{\text{th}}(x,y)+k_{c}^{3}w_{\text{ath}}(\bar{x},\bar{y})%
\genfrac{}{}{0pt}{1}{{}}{{}}%
\right)  ~,\label{IV TF n D}%
\end{align}
with the following notation for the derivatives:%
\begin{align}
u_{\text{th}}(x,y) &  =\frac{\partial f_{\text{th}}(x,y)}{\partial x}~,\quad
u_{\text{ath}}(\bar{x},\bar{y})=\frac{\partial f_{\text{ath}}(\bar{x},\bar
{y})}{\partial\bar{x}}~,\nonumber\\
w_{\text{th}}(x,y) &  =-\frac{\partial f_{\text{th}}(x,y)}{\partial y}~,\quad
w_{\text{ath}}(\bar{x},\bar{y})=-\frac{\partial f_{\text{ath}}(\bar{x},\bar
{y})}{\partial\bar{y}}~.\label{IV TF f derivs}%
\end{align}
The entropy has a contribution only from the thermal part, and is given by
\begin{equation}
S_{\mathrm{ni}}\simeq k_{\text{B}}\int\frac{d\mathbf{r}}{\lambda_{T}^{3}%
}\left(  -\frac{5}{2}f_{\text{th}}\left(  x,y\right)  +xu_{\text{th}}\left(
x,y\right)  -yw_{\text{th}}(x,y)\right)  ~.\label{IV S TF}%
\end{equation}
The Hohenberg--Kohn free energy becomes
\begin{multline}
F_{\text{nc}}\left(  \left[  n_{\text{nc}},\Delta_{\text{nc}},\Delta
_{\text{nc}}^{\ast}\right]  ,T\right)  \simeq\label{IV F TF}\\
\int d\mathbf{r}\left(  \frac{k_{\text{B}}T}{\lambda_{T}^{3}}\tilde
{f}_{\text{th}}\left(  u_{\text{th}},w_{\text{th}}\right)  +\mathcal{E}%
_{c}k_{c}^{3}\tilde{f}_{\text{ath}}\left(  u_{\text{ath}},w_{\text{ath}%
}\right)  \right)  ~,
\end{multline}
where the Legendre transforms are defined as [again, the minimization may be
replaced by the requirements of Eq.~(\ref{IV TF f derivs})]
\begin{align}
\tilde{f}_{\text{th}}\left(  u_{\text{th}},w_{\text{th}}\right)   &
=\max_{x,y}\left(  f_{\text{th}}\left(  x,y\right)  -xu_{\text{th}%
}+yw_{\text{th}}%
\genfrac{}{}{0pt}{1}{{}}{{}}%
\right)  ~,\nonumber\\
\tilde{f}_{\text{ath}}\left(  u_{\text{ath}},w_{\text{ath}}\right)   &
=\max_{\bar{x},\bar{y}}\left(  f_{\text{ath}}\left(  \bar{x},\bar{y}\right)
-xu_{\text{ath}}+yw_{\text{ath}}%
\genfrac{}{}{0pt}{1}{{}}{{}}%
\right)  ~,\label{IV tilde f}%
\end{align}
and are used with the conditions%
\begin{align}
\lambda_{T}^{3}u_{\text{th}}+k_{c}^{3}u_{\text{ath}} &  =n_{\text{nc}}%
~,\quad2\lambda_{T}^{3}w_{\text{th}}+2k_{c}^{3}w_{\text{ath}}=\left\vert
\Delta_{\text{nc}}\right\vert ~,\nonumber\\
k_{\text{B}}T\frac{\partial\tilde{f}_{\text{th}}}{\partial u_{\text{th}}} &
=\mathcal{E}_{c}\frac{\partial\tilde{f}_{\text{ath}}}{\partial u_{\text{ath}}%
}~,\quad k_{\text{B}}T\frac{\partial\tilde{f}_{\text{th}}}{\partial
w_{\text{th}}}=\mathcal{E}_{c}\frac{\partial\tilde{f}_{\text{ath}}}{\partial
w_{\text{ath}}}~.\label{IV Leg conds}%
\end{align}
Further details regarding $f_{\text{th}}$\ and $f_{\text{ath}}$\ are given
in the appendix, and for  $f_{\text{th}}$\ also in Sec. V.

As for the $\Phi$-DFT of the previous section, a local, Thomas--Fermi approach
can be applied to the condensate amplitude $\Phi$ as well, amounting to
neglecting the derivative term in Eq.~(\ref{IV phi cond}), resulting in a
completely local description.

\subsection{Interaction Effects}

A local description of the interaction effects (LDA) for the present
application of DFT requires knowledge of $f_{\text{int}}(n,\Phi,\Phi^{\ast
},\Delta,\Delta^{\ast},T)$ --- the interaction energy density for a uniform
system of density $n$, condensate amplitude $\Phi$, and anomalous density
$\Delta$, at temperature $T$. Its derivatives will determine the potentials of
the reference system, which include
\begin{equation}
\xi_{\mathrm{ni}}(\mathbf{r})=\xi_{\text{ext}}(\mathbf{r})+\xi_{\text{int}%
}(\mathbf{r})=-{\frac{\partial f_{\text{int}}}{\partial\Delta^{\ast}}}\,,
\label{IV xieff}%
\end{equation}
in addition to Eq.~(\ref{III LDA pots}) for the interaction contribution to
the effective potential $v_{\mathrm{ni}}$ and fictitious field $\eta
_{\mathrm{ni}}$.

As in the previous section, one may use the weakness of the interactions in
order to derive a simple approximation for $f_{\text{int}}$, by evaluating the
interaction energy to leading order in the interaction strength $g$.
\begin{align}
f_{\text{int}}  &  \simeq{\frac{g}{2}}\left(  |\Phi|^{4}+4|\Phi|^{2}%
\langle\hat{\phi}^{\dagger}\hat{\phi}\rangle+\right. \label{IV f int}\\
&  \qquad\left.  \Phi^{\ast2}\langle\hat{\phi}\hat{\phi}\rangle+\Phi
^{2}\langle\hat{\phi}^{\dagger}\hat{\phi}^{\dagger}\rangle+\langle\hat{\phi
}^{\dagger}\hat{\phi}^{\dagger}\hat{\phi}\hat{\phi}\rangle\right)
\;,\nonumber
\end{align}
due to the mixing of creation and annihilation operators in
Eq.~(\ref{IV gammas}), since expectation values of the type $\langle\hat{\phi
}\hat{\phi}\rangle$ no longer vanish. Similarly, Wick's theorem becomes
$\langle\hat{\phi}_{1}^{\dagger}\hat{\phi}_{2}^{\dagger}\hat{\phi}_{3}%
\hat{\phi}_{4}\rangle=\langle\hat{\phi}_{1}^{\dagger}\hat{\phi}_{2}^{\dagger
}\rangle\langle\hat{\phi}_{3}\hat{\phi}_{4}\rangle+\langle\hat{\phi}%
_{1}^{\dagger}\hat{\phi}_{4}\rangle\langle\hat{\phi}_{2}^{\dagger}\hat{\phi
}_{3}\rangle+\langle\hat{\phi}_{1}^{\dagger}\hat{\phi}_{3}\rangle\langle
\hat{\phi}_{2}^{\dagger}\hat{\phi}_{4}\rangle$. Using both $n=|\Phi
|^{2}+\langle\hat{\phi}^{\dagger}\hat{\phi}\rangle$, and $\Delta=\Phi
^{2}+\langle\hat{\phi}\hat{\phi}\rangle$ gives
\begin{equation}
f_{\text{int}}\simeq{\frac{g}{2}}\left(  2n^{2}+|\Delta|^{2}-2|\Phi
|^{4}\right)  \;. \label{IV f int f}%
\end{equation}
This result can be interpreted as a triple counting of the condensate term,
compensated by a double subtraction --- the $|\Phi|^{4}$ contribution appears
in the direct ($\langle\hat{\psi}_{1}^{\dagger}\hat{\psi}_{4}\rangle
\langle\hat{\psi}_{2}^{\dagger}\hat{\psi}_{3}\rangle$), the exchange
($\langle\hat{\psi}_{1}^{\dagger}\hat{\psi}_{3}\rangle\langle\hat{\psi}%
_{2}^{\dagger}\hat{\psi}_{4}\rangle$) and the anomalous ($\langle\hat{\psi
}_{1}^{\dagger}\hat{\psi}_{2}^{\dagger}\rangle\langle\hat{\psi}_{3}\hat{\psi
}_{4}\rangle$) term, but physically should be accounted for only once. Using
this approximation in Eqs.~(\ref{III LDA pots}) and (\ref{IV xieff}) for the
effective potentials gives
\begin{align}
v_{\mathrm{ni}}(\mathbf{r})  &  \simeq v_{\text{ext}}(\mathbf{r}%
)+2gn(\mathbf{r})\text{\ ,}\nonumber\\
\eta_{\mathrm{ni}}(\mathbf{r})  &  \simeq2g|\Phi(\mathbf{r})|^{2}%
\Phi(\mathbf{r})\text{\ ,}\label{IV LDA pots}\\
\xi_{\mathrm{ni}}(\mathbf{r})  &  \simeq-{\frac{g}{2}}\Delta(\mathbf{r}%
)\text{\ .}\nonumber
\end{align}

Substitution of this approximation into Eq.~(\ref{IV phi cond}) gives the
generalized Gross--Pitaevskii equation
\begin{equation}
\left(  \! \! -\frac{\hbar^{2}\nabla^{2}}{2m}+v_{\text{ext}}-\mu
+2gn_{\text{nc} }+g|\Phi(\mathbf{r})|^{2} \! \! \right)  \Phi+g\Delta
_{\text{nc}}\Phi^{\ast}=0 ~, \label{IV GP}%
\end{equation}
which contains an extra term involving $\Delta_{\text{nc}}$
[cf.~Eq.~(\ref{III GP})].  These potentials are also to be used in the
Bogoliubov--de Gennes equation, Eq.~(\ref{IV BdG}).

This approach reproduces the equations of the
Hartree--Fock--Bogoliubov method, which have been systematically
derived and studied for a long time \cite{Dalfovo,Griffin96,
HuangYang, STB58, Popov,BlaizotRipka}.  In that context, the method is
intended to calculate not only the thermodynamic properties of the
system, but also its excitation spectrum.  It has been criticized for
producing a spectrum with a gap at small wavenumbers in homogeneous
systems, while it is known that the correct long--wavelength result
involves a gapless, linear phonon spectrum \cite{Griffin96}.  A
related difficulty arises at short distances, where it is seen that
$\Delta(\mathbf{r}) =\lim_{\mathbf{r}^{\prime }\rightarrow\mathbf{r}}
\langle\hat{\psi}(\mathbf{r}) \hat{\psi}(\mathbf{r}^{\prime})
\rangle$ depends on the long--wavenumber cutoff $k_c$ (for $T=0$
and weak interactions, one may use the limit of $f_{\text{ath}}$
discussed in the appendix to find $n_{\text{nc}}/n\sim
\sqrt{na_{0}^{3}}$ and $\Delta_{\text{nc}} / n \sim k_c a_{0}$, in
agreement with the literature).  As $\Delta_{\text{nc}}$ involves a
product of operators at essentially the same point in space, such a
cutoff dependence should be accepted, and indeed, the nature of the
point interaction should be expected to generate a relationship
between this ultraviolet divergence and long--wavelength behavior.
Approximations which are designed to produce a spectrum without a gap
have been studied \cite{Popov,STB58}, and more recently, a
pseudopotential which allows for a rigorous treatment overcoming these
difficulties (at least at zero temperature) has been suggested
\cite{Olshanii}.

In the context of DFT, it may be argued that the spectrum is irrelevant, as it
represents a property of the reference system which need not be shared with
the interacting system. In principle, it may thus be claimed that A-DFT is
rigorously exact, despite the presence of the gap. However, it is clear that
the spectrum affects the thermodynamic properties, and that if the reference
system has properties which differ significantly from those of the interacting
system, it will be difficult to find workable approximations for
$F_{\text{int}}$. It may thus be desirable to adopt the advanced
pseudopotential approach to DFT \cite{Olshanii}. This would require using a
reference system with nonlocal effective potentials, i.e., introducing terms
of the form $\xi( \mathbf{r,r}^{\prime}) \psi^{\dag}( \mathbf{r}) \psi^{\dag}(
\mathbf{r}^{\prime})$ and $\upsilon( \mathbf{r,r}^{\prime}) \psi^{\dag}(
\mathbf{r}) \psi( \mathbf{r}^{\prime})$ into the Hamiltonian, with
$\xi_{\text{ext}}=\upsilon_{\text{ext}}=0$ but $\xi_{\mathrm{ni}}%
,\upsilon_{\mathrm{ni}}\neq0$ in close analogy to the derivation above, and is
beyond the scope of the present work.

\section{The uniform--system, weak--interactions limit}

Having introduced both $\Phi$-DFT and A-DFT, a discussion of applications is
in order. Essentially all applications are beyond the scope of the present
article, but one ``application'' which is particularly revealing will be
presented here: the weak--interactions limit of homogeneous systems. Strictly
speaking, this is not an application of DFT, and should be viewed instead as
an exercise in thermodynamic perturbation theory. It will have relevance to
the local density approximation, which (as noted above) relies on knowledge of
the properties of uniform systems.

The Thomas--Fermi approach is accurate (not an approximation) for homogeneous
systems. Hence, the results of the Thomas--Fermi subsections above are
directly applicable. Similarly, the lowest order, linear term in the
interaction strength $g$ has been given above, for both $\Phi$-DFT and A-DFT.
The discussion here pertains to a finite temperature, not too near to either
the BEC transition or to $T=0$, and thus, the interactions are treated as
perturbing a noninteracting Bose--condensed system.

\subsection{Two--fluid method}

Applying the $\Phi$-DFT of Sec.~III to a uniform system, one drops the
gradient terms in the Gross--Pitaevskii equation (\ref{III GP}), and solves it
in conjunction with Eq.~(\ref{III TF n f'}), in the $g\rightarrow0$ limit. The
external potential is set to $0$, and the $x\rightarrow0$ limit of the
thermodynamic function $f$ defined in Eq.~(\ref{III f(x)}) is pertinent (cf.
Fig. \ref{Fig fff}):
\begin{equation}
f(x)=-\zeta(5/2)+\zeta(3/2) x -{\frac{4\sqrt{\pi}}{3}}x^{3/2}+O(x^{5/2}) ~,
\label{V 2f f}%
\end{equation}
giving
\begin{equation}
u=f^{\prime}(x) = \zeta({3/2}) - 2\sqrt{\pi}x^{1/2}+O(x^{3/2})~,
\label{V 2f f'}%
\end{equation}
where $x=\left(  2gn-\mu\right)  /k_{\text{B}}T$ from Eq.
(\ref{III 1st O eff pots}) and $u=\lambda_{T}^{3}n_{\text{th}}$. This relation
between the effective potential and the density may be inverted as
\begin{equation}
x=\frac{\delta u^{2}}{4\pi}+O\left(  \delta u^{4}\right)  ~, \label{V 2f x}%
\end{equation}
where the notation $\delta u=\zeta(3/2) - u$ has been used, and the Legendre
transformed Helmholtz free energy is, in dimensionless form,
\begin{equation}
\tilde{f}(u)=-\zeta(5/2) + \frac{\delta u^{3}}{12\pi} + O(\delta u^{5}) ~.
\label{V 2f tilde f}%
\end{equation}

The Hohenberg--Kohn free energy per unit volume of the uniform system,
including the interaction terms to leading order in $g$ from Eq.
(\ref{III Fint 1st O}), is thus:
\begin{multline}
{\frac{F_{\text{HK}}(\left[  n,\Phi,\Phi^{\ast} \right]  ,T)}{V}}\simeq
\frac{k_{\text{B}}T}{\lambda_{T}^{3}} \tilde{f}( \lambda_{T}^{3} (n-\Phi
^{\ast}\Phi) )\label{V 2f F}\\
+{\frac{g}{2}}\left(  2n^{2}-\left(  \Phi^{\ast}\Phi\right)  ^{2} \right)  ~.
\end{multline}
Although it is straightforward to apply the self--consistent
Kohn--Sham equations to this system, which amounts here to using,
e.g., Eq.~(\ref{V 2f f'}), a more physically transparent discussion
will result from following a minimum--energy path, closer in spirit to
the Hohenberg--Kohn approach.  The physical requirement that the
external auxiliary field vanish, $\eta_{\text{ext}}=\partial
F_{\text{HK}}/\partial\Phi^{\ast}=0$, will thus be imposed by
minimizing the free energy of Eq.~(\ref{V 2f F}) with respect to
$\Phi^{\ast}$, at a given overall density $n$.  From Eq.~(\ref{V 2f
tilde f}), this amounts to minimization of
\begin{equation}
\frac{\delta u^{3}}{6\pi}-\frac{g}{k_{\text{B}}T\lambda_{T}^{3}}\left(
\lambda_{T}^{3}n-\zeta(3/2) + \delta u\right)  ^{2} \label{V 2f to min}%
\end{equation}
with respect to $\delta u=\lambda_{T}^{3}(|\Phi|^{2}-n) + \zeta(3/2)$, with
$\delta u\geq0$, but small. At $g=0$, one finds $\delta u=0$ or $\left\vert
\Phi_{0}\right\vert ^{2}=n-\zeta(3/2) \lambda_{T}^{-3}$, whereas for small $g$
one finds $\delta u\simeq2\sqrt{{\frac{\pi g}{k_{\text{B}}T}}\left\vert
\Phi_{0}\right\vert ^{2}}$. This gives
\begin{equation}
|\Phi|^{2}\simeq|\Phi_{0}|^{2}+2\lambda_{T}^{-3}\sqrt{{\frac{\pi
g}{k_{\text{B}}T}}|\Phi_{0}|^{2}}\text{\ ,} \label{V 2f Phi}%
\end{equation}
a result whose accuracy should be questioned, as discussed below. Note that
the particularly soft $\delta u^{3}$ behavior of $\tilde{f}\left(  u\right)
$, corresponding to the nonanalytic cusp in $f\left(  x\right)  $ at $x=0$,
has resulted in a sensitivity to interactions which is displayed by the sharp
$\sqrt{g}$ dependence of the condensate amplitude (in other words, the
position of a shallow minimum is easily changed by a perturbation which is
sloped in that region). It is straightforward but not particularly
illuminating to obtain additional thermodynamic results at this level of
approximation, e.g., to find the relationship between the chemical potential
$\mu$ and the overall density $n$.

\subsection{Bogoliubov method}

Application of the A-DFT of Sec. IV to the present problem requires expanding
the thermodynamic functions defined in Eq.~(\ref{IV fT fA}) at small values of
their arguments [see the appendix, Eqs.~(\ref{A I1}) and (\ref{A I2})]:%
\begin{multline}
f_{\text{th}}(x,y)=-\zeta(5/2)+\zeta(3/2)x\label{V B fT}\\
-{\frac{2\sqrt{\pi}}{3}}\left(  \left(  x+2y\right)  ^{3/2}+\left(
x-2y\right)  ^{3/2}\right)  +O(x^{2})~,
\end{multline}
where the requirement $0\leq y\leq x/2$ is used to drop $O\left(
y^{2}\right)  $\ contributions. A comparison with Eq.~(\ref{V 2f f}) is
interesting already at this stage: clearly, setting the anomalous potential to
zero, $y=0$, reproduces the $\Phi$-DFT result, but taking the Bogoliubov
spectrum with a vanishing gap corresponds here to setting $y=x/2$, and results
in an extra factor of $\sqrt{2}$\ in the $x^{3/2}$ term. The expansion of
$f_{\text{ath}}$ begins with $O(x^{2})$ terms, Eq.~(\ref{A fA}), and therefore
the athermal component does not affect the results to leading order in $g$.

The scaled density and anomalous density are%
\begin{equation}
u_{\text{th}}=\frac{\partial f_{\text{th}}}{\partial x}\simeq\zeta(3/2)
-\sqrt{\pi}\left(  \sqrt{x+2y}+\sqrt{x-2y}\right)  \label{V B u}%
\end{equation}
and
\begin{equation}
w_{\text{th}}=-\frac{\partial f_{\text{th}}}{\partial x}\simeq2\sqrt{\pi
}\left(  \sqrt{x+2y}-\sqrt{x-2y}\right)  ~, \label{V B w}%
\end{equation}
and inverting these relationships gives%
\begin{equation}
x\simeq\frac{1}{4\pi}\left(  \delta u_{\text{th}}^{2}+\frac{1}{4}w_{\text{th}%
}^{2}\right)  ~, \quad y\simeq\frac{1}{8\pi}\delta u_{\text{th}}w_{\text{th}%
}~, \label{V B xy}%
\end{equation}
where again $\delta u_{\text{th}}=\zeta(3/2) - u_{\text{th}}$. The result for
the Hohenberg--Kohn free energy is
\begin{equation}
\tilde{f}_{\text{th}}(u_{\text{th}},w_{\text{th}}) = -\zeta(5/2) + {\frac
{1}{4\pi}} ( {\frac{1}{3} }\delta u_{\text{th}}^{3}+{\frac{1}{4}}w_{\text{th}%
}^{2}\delta u_{\text{th}}) +O(\delta u_{\text{th}}^{4}) ~.
\label{V B tilde fT}%
\end{equation}

The overall free energy, including the interaction terms to leading order, Eq.
(\ref{IV f int f}), but excluding the athermal contribution, is:
\begin{align}
&  {\frac{F_{\mathrm{HK}}(n,\Phi,\Phi^{\ast},\Delta,\Delta^{\ast},T)}{V}%
}\simeq\label{V B F HK}\\
&  \quad k_{\text{B}}T\lambda_{T}^{-3}\tilde{f}_{\text{th}}\left(  \lambda
_{T}^{3}(n-|\Phi|^{2}),\lambda_{T}^{3}\left\vert \Delta-\Phi^{2}\right\vert
\genfrac{}{}{0pt}{1}{{}}{{}}%
\right)  +\nonumber\\
&  \qquad\qquad\qquad\qquad\qquad\qquad\qquad{\frac{g}{2}}(2n^{2}+|\Delta
|^{2}-2|\Phi|^{4})~.\nonumber
\end{align}
In the present case, both the fictitious potentials $\eta$ and $\xi$ must
vanish, i.e., the free energy is to be minimized with respect to both $\Phi$
and $\Delta$. In dimensionless terms, this corresponds to minimization of
\begin{multline}
{\frac{1}{2\pi}}\left(  {\frac{1}{3}}\delta u_{\text{th}}^{3}+{\frac{1}{4}%
}\delta u_{\text{th}}w_{\text{th}}^{2}\right)  +{\frac{g}{k_{\text{B}}%
T\lambda_{T}^{3}}}\times\\
\left(  \left(  \lambda_{T}^{3}n-\zeta(3/2)+\delta u_{\text{th}}%
-\frac{w_{\text{th}}}{2}\right)  ^{2}\right.  \\
\left.  -2\left(  \lambda_{T}^{3}n-\zeta(3/2)+\delta u_{\text{th}}\right)
^{2}\right)  \label{V B min}%
\end{multline}
with respect to both $\delta u_{\text{th}}$ and $w_{\text{th}}$ (in principle,
an arbitrary phase factor could be associated with the $\frac{w_{\text{th}}%
}{2}$ term, reflecting the relative phase between $\Delta_{\text{nc}}$\ and
$\Phi^{2}$, but the sign used here is clearly optimal). \ The minimization
requires
\begin{align}
\delta u_{\text{th}}^{2}+{\frac{1}{4}}w_{\text{th}}^{2} &  \simeq{\frac{4\pi
g}{k_{\text{B}}T\lambda_{T}^{3}}}\left(  \lambda_{T}^{3}n-\zeta(3/2)+\delta
u_{\text{th}}+\frac{w_{\text{th}}}{2}\right)  ~,\nonumber\\
\delta u_{\text{th}}w_{\text{th}} &  \simeq{\frac{16\pi g}{k_{\text{B}%
}T\lambda_{T}^{3}}}\left(  \lambda_{T}^{3}n-\zeta(3/2)+\delta u_{\text{th}%
}-\frac{w_{\text{th}}}{2}\right)  ~.\label{V B =0}%
\end{align}
To leading order in $g$ we may use $\delta u_{\text{th}}\simeq0$ and
$w_{\text{th}}\simeq0$ in the RHS, resulting in $\delta u_{\text{th}}\simeq$
$w_{\text{th}}/2\simeq\sqrt{{\frac{2\pi g}{k_{\text{B}}T\lambda_{T}^{3}}%
}\left(  \lambda_{T}^{3}n-\zeta(3/2)\right)  }$, or
\begin{equation}
|\Phi|^{2}\simeq|\Phi_{0}|^{2}+\lambda_{T}^{-3}\sqrt{{\frac{2\pi
g}{k_{\text{B}}T}}|\Phi_{0}|^{2}}\text{\ ,}\label{V B Phi}%
\end{equation}
together with a similar contribution to $\Delta_{\text{nc}}$ (at this order,
we have $x=2y$). The interaction--dependent correction to the condensate
fraction is here a factor of $\sqrt{2}$ smaller than the result of the
previous subsection, Eq.~(\ref{V 2f Phi}). This discrepancy will be discussed
in the next subsection. The present result, which includes the effects of the
anomalous density through $w$, is in accordance with the literature
\cite{Popov}\ (recall that $g\simeq4\pi\hbar^{2}a_{0}/m$ and $\lambda
_{T}=\sqrt{2\pi\hbar^{2}/mk_{\text{B}}T}$).

\subsection{Comparison of the two--fluid and Bogoliubov Reference System}

It is at first surprising that the two methods discussed above lead to two
different results for the leading--order correction to the condensate fraction
of a uniform BEC. After all, both methods are based on a straightforward
expansion in the small parameter $g$, within a thermodynamic framework which
is in principle exact. The only source of error can be the neglect of terms in
the interaction energy, $f_{\text{int}}$, which are of higher order in $g$.
The next paragraphs display these terms explicitly.

Clearly, the level of accuracy used in the A-DFT description above can be
imported into $\Phi$-DFT, by choosing the appropriate form for $f_{\text{int}%
}\left(  n,\Phi,\Phi^{\ast}\right)  $. In fact, it is obvious that simply
dropping the anomalous term in Eq.~(\ref{V B min}), setting $w_{\text{th}}=0$,
reduces it to Eq.~(\ref{V 2f to min}). However, the proper procedure is to
minimize over $w_{\text{th}}$, which corresponds to imposing the condition
$\xi=0$, i.e., to requiring the vanishing of the anomalous potential rather
than the $w_{\text{th}}$ contribution to the anomalous density. From the
second line in Eq. (\ref{V B =0}), this leads (to leading order in $g$) to%
\begin{equation}
w_{\text{th}}\simeq{\frac{16\pi g}{k_{\text{B}}T\lambda_{T}^{3}}}\frac
{\lambda_{T}^{3}n-\zeta(3/2)}{\delta u_{\text{th}}}~.\label{V c w}%
\end{equation}
Introducing this into Eq.~(\ref{V B min}) gives%
\begin{multline}
{\frac{1}{6\pi}}\delta u_{\text{th}}^{3}-{\frac{g}{k_{\text{B}}T\lambda
_{T}^{3}}}\left(  \lambda_{T}^{3}n-\zeta(3/2)+\delta u_{\text{th}}\right)
^{2}\label{V c to min}\\
+16\pi\left(  {\frac{g}{k_{\text{B}}T\lambda_{T}^{3}}}\right)  ^{2}%
\frac{\left(  \lambda_{T}^{3}n-\zeta_{3/2}\right)  ^{2}}{\delta u_{\text{th}}}%
\end{multline}
where terms up to second order in $g$ have been retained (one can check
\textit{a posteriori} that the neglected terms are indeed small). Minimization
of this, to leading order, amounts to the requirement that
\begin{equation}
\left(  \delta u_{\text{th}}^{2}-{\frac{2\pi g}{k_{\text{B}}T\lambda_{T}^{3}}%
}\left(  \lambda_{T}^{3}n-\zeta(3/2)\right)  \right)  ^{2}=0~,\label{V c =0}%
\end{equation}
which displays how dropping terms of second order in $g$ leads to a doubling
of the result for $\delta u_{\text{th}}^{2}$. It is thus clarified that a
weakly--interacting Bose--Einstein condensed system is situated near a
singular point, associated physically with a \textquotedblleft completely
full\textquotedblright\ thermal cloud, $u_{\text{th}}=\zeta(3/2)$. At this
special point, terms which are of second order in the weak interaction
parameter $g$ are not relatively small, because they are divergent, with the
small quantity $\delta u_{\text{th}}$ appearing in the denominator.

\section{Summary and outlook} \label{Sec:summary}

The thermodynamic approach (summarized in Sec.  II) provides a general
method for generating DFTs for bosonic systems in thermal equilibrium
at finite temperature, and has been used to derive the
equations of $\Phi$-DFT (Sec.  III) and of A-DFT (Sec.  IV).  The
different DFTs use as references different types of Kohn--Sham
systems, which are subject to different fictitious potentials.  The
reference systems have quadratic, soluble Hamiltonians, and the
interaction effects are to be included via a local--density
approximation.  The latter must be based on knowledge of the free
energy of interacting homogeneous systems, which are 
subject to the fictitious potentials.  For $\Phi$-DFT, knowledge of
the free energy of a homogeneous interacting system as a function of
$\mu$, $T$ and $\eta$ is required in order to supply
$f_{\text{int}}\left( n,\Phi,\Phi^{\ast },T\right) $, and for A-DFT a
$\xi$ field must also be allowed for.  This type of information is
generally available only to leading order in the interaction parameter
$g$.  In this limit, our results for $\Phi$-DFT generalize the
two--fluid approach \cite{MCT97} to inhomogeneous systems, and the
results for A-DFT reproduce the Hartree--Fock--Bogoliubov model.  As
has occurred for electronic systems, we anticipate that the necessary
results beyond the leading order will be generated using quantum Monte
Carlo techniques.  Such techniques have been developed for
Bose--condensed systems \cite{Krauth}, but as the fictitious
potentials $\eta$ and $\xi$ break particle--number conservation,
different variants of the techniques may need to be developed to meet
this goal.

It is of interest to compare $\Phi$-DFT and A-DFT to the attempt to
apply DFT to Bose--condensed systems made in Ref.~\cite{AG95}, using
the standard approach to DFT rather than the thermodynamic one.  In
this reference, only two functional variables were used --- $n$ and
$\Phi$, as in $\Phi$-DFT --- but the reference or Kohn--Sham system
chosen employed a Bogoliubov type treatment, similar to that used in
the A-DFT above.  Correspondingly, the Hamiltonian of the reference
system depends not only on the potentials used, but also on the
(anomalous) density, which is to be calculated self--consistently.
This represents a difficulty which the present thermodynamic
derivation avoids.

In comparing the two DFT methods, it was found that in the limit of
homogeneous, weakly--interacting systems at finite temperatures, $\Phi$-DFT
(or the two--fluid method) does not correctly reproduce the leading--order
correction to the condensate density, and that this flaw can be corrected by
including higher--order terms in $f_{\text{int}}$. This is due to the extreme
sensitivity of the corresponding energy--minimization problem: (a) terms of
order $g$ in the energy cause a shift in the minimizing value of $\Phi$ which
is of order $\sqrt{g}$, and (b) the second order term in $f_{\text{int}}$ is
divergent, having an $O\left(  \sqrt{g}\right)  $ denominator, thus
contributing to $O\left(  g^{3/2}\right)  $ in the energy and to a significant
change in the $O\left(  \sqrt{g}\right)  $ correction to $\Phi$. The
higher--order terms in $f_{\text{int}}$ for $\Phi$-DFT were in this case
obtained from a first--order A-DFT calculation, which amounts to
re--expressing the Hartree--Fock--Bogoliubov model as a minimization problem
(Sec. V).

The comparison just mentioned demonstrates the advantages of A-DFT -- it uses
a reference system which is much closer in its behavior to the fully
interacting system, and therefore the approximation introduced is much less
significant.  It is reasonable to expect that this advantage will be
significant for inhomogeneous systems as well. Thus, obtaining the requisite
data for homogeneous interacting systems as a function of both $\eta$ and
$\xi$ is called for.  As an interim step, applications of $\Phi$-DFT, for
which strong--interaction corrections to $f_{\text{int}}$ could more easily be
acquired, should also be considered.

\begin{acknowledgments}
The authors wish to thank Y. Avishai for the role he played in initial stages
of this project. This work was supported in part by grants from the
U.S.-Israel Binational Science Foundation (No.~2006212), the Israel Science
Foundation (No.~29/07), and the James Franck German-Israel Binational Program.
\end{acknowledgments}

\appendix

\section{Integrals for the Thomas--Fermi approximation of A-DFT}

In Sec.~IV, Eq.~(\ref{IV fT fA}), integrals corresponding to the thermodynamic
functions $f_{\text{th}}$ and $f_{\text{ath}}$ were introduced, and in Sec. V,
the need to evaluate these integrals in the limit corresponding to weak
interactions arose. The details of the evaluations are presented here.

For the thermal contribution, integration of Eq. (\ref{IV fT}) by parts gives
\begin{align}
f_{\text{th}}\left( x,y\right) &
={\frac{-8}{3\sqrt{\pi}}}\int_{0}^{\infty}
{\frac{\sqrt{(q^{2}+x)^{2}-4y^{2}}}{\exp(\sqrt{(q^{2}+x)^{2}-4y^{2}
})-1}}\times \nonumber \\
&  {\frac{q^{4}(q^{2}+x)dq}{(q^{2}+x)^{2}-4y^{2}}}~.
\end{align}
The second factor is separated as
\begin{align}
&  {\frac{q^{4}(q^{2}+x)}{(q^{2}+x)^{2}-4y^{2}}}=q^{2}-x+ \nonumber\\
& {\frac{1}{2}}\left( {\frac{(x-2y)^{2}}{q^{2}+x-2y}} +
{\frac{(x+2y)^{2} }{q^{2}+x+2y}}\right) ~,
\end{align}
where the term $q^{2}-x$ diverges with $q$, and the remaining terms converge
rapidly. This may be used to write $f_{\text{th}}=$ $I_{1}+I_{2}$, with the
integrals $I_{1}$ and $I_{2}$ involving the divergent and convergent terms, respectively.

At small $x$ and $y$, the square root term can be expanded as
\begin{equation}
\sqrt{(q^{2}+x)^{2}-4y^{2}}=q^{2}+x+O(x^{2})\;,
\end{equation}
where terms of order $y^{2}$ are included in $O(x^{2})$ because of the
limitation $x\geq2y\geq0$. One finds%
\begin{align}
I_{1} &  ={\frac{-8}{3\sqrt{\pi}}}\int_{0}^{\infty}{\frac{q^{2}+x}{\exp
(q^{2}+x)-1}}(q^{2}-x)dq+O(x^{2})\nonumber\label{A I1}\\
&  =-\zeta(5/2)+\zeta(3/2)x+O(x^{2})
\end{align}
(note cancellation of terms at order $x^{3/2}$) and
\begin{align}
I_{2} &  ={\frac{-4}{3\sqrt{\pi}}}\int_{0}^{\infty}\left(  {\frac{(x-2y)^{2}%
}{q^{2}+x-2y}}+{\frac{(x+2y)^{2}}{q^{2}+x+2y}}\right)  dq+O(x^{2}%
)\nonumber\label{A I2}\\
&  ={\frac{-2\sqrt{\pi}}{3}}\left(  (x-2y)^{3/2}+(x+2y)^{3/2}\right)
+O(x^{2})~,
\end{align}
where in the last equation $(q^{2}+x)/($e$^{q^{2}+x}-1)$ was approximated by
unity because the remaining factors are already of relatively high order in
$x$, and are small when $q^{2}\gg x$. \ These results are used in Sec. V, Eq.
(\ref{V B fT}).

For the athermal contribution, one may rewrite Eq. (\ref{IV fA}) as%
\begin{multline}
f_{\text{ath}}\left(  \bar{x},\bar{y}\right)  =-\frac{4}{\sqrt{\pi}}\bar
{y}^{2}+\\
{\frac{2}{\sqrt{\pi}}}\int_{0}^{1}\left[  \left(  \sqrt{(q^{2}+\bar{x}%
)^{2}-4\bar{y}^{2}}-q^{2}-\bar{x}\right)  q^{2}+2\bar{y}^{2}\right]  dq~,
\end{multline}
where the integral can be continued to infinity without divergence, as can be
seen by expanding the square root as
\begin{equation}
\sqrt{(q^{2}+\bar{x})^{2}-4\bar{y}^{2}}=q^{2}+\bar{x}-\frac{2\bar{y}^{2}%
}{q^{2}}+O\left(  \bar{x}^{3}\right)  ~.
\end{equation}
Rescaling $q$ by $\sqrt{\bar{x}}$, and defining
\begin{equation}
h\left(  \alpha\right)  =\int_{0}^{\infty}\left[  \left(  \sqrt{\left(
q^{2}+1\right)  ^{2}-\alpha}-q^{2}-1\right)  q^{2}+\frac{\alpha}{2}\right]
dq~,\label{A h}%
\end{equation}
where $\alpha=4\bar{y}^{2}/\bar{x}^{2}$ is a variable in the range
$0\leq\alpha\leq1$, gives
\begin{equation}
f_{\text{ath}}\left(  \bar{x},\bar{y}\right)  =-{\frac{4}{\sqrt{\pi}}}\bar
{y}^{2}+{\frac{2\bar{x}^{5/2}}{\sqrt{\pi}}}h(\alpha)+O\left(  \bar{x}%
^{3}\right)  ~.\label{A fA}%
\end{equation}
The integral $h\left(  \alpha\right)  $ vanishes at $\alpha=0$, and increases
with $\alpha$ to a value of $8\sqrt{2}/15$ at $\alpha=1$. This value may be
derived as
\begin{multline}
\int_{0}^{\infty}\left[  \left(  \sqrt{q^{4}+2q^{2}}-q^{2}-1\right)
q^{2}+\frac{1}{2}\right]  dq=\\
\lim_{Q\rightarrow\infty}-\frac{Q^{5}}{5}-\frac{Q^{3}}{3}+\frac{Q}{2}+\int%
_{2}^{Q^{2}+2}\sqrt{u}\left(  u-2\right)  \frac{du}{2}~,
\end{multline}
where $u=q^{2}+2$ and the result obtains from the lower limit of the last
integral. The derivative of $h$ is given by
\begin{equation}
h^{\prime}(\alpha)=\frac{1}{2}\int_{0}^{\infty}\left(  1-\frac{q^{2}}%
{\sqrt{\left(  q^{2}+1\right)  ^{2}-\alpha}}\right)  dq~,
\end{equation}
and decreases from $h^{\prime}(0)=\pi/4$ to $h^{\prime}(1)=1/\sqrt{2}$. As
these values are within 10\% of each other, a plot of $h$ is very nearly a
straight line, and is not included. It is easily seen that $h^{\prime}\left(
\alpha\right)  =\pi/4-\left(  \pi/64\right)  \alpha+O\left(  \alpha
^{2}\right)  $ for small $\alpha$, whereas $h^{\prime}\left(  \alpha\right)
=1/\sqrt{2}-\left(  1/16\sqrt{2}\right)  \left(  1-\alpha\right)  \log\left(
1-\alpha\right)  +O\left(  1-\alpha\right)  $ near $\alpha=1$, where the
logarithm arises from the $1/q$ behavior of the integrand for $\left(
1-\alpha\right)  \ll q^{2}\ll1$. This motivates the approximation
\begin{equation}
h(\alpha)\simeq c_{1}\alpha+c_{2}\alpha^{2}+c_{3}\alpha^{3}+c_{4}\alpha
^{4}+\check{c}\left(  1-\alpha\right)  ^{2}\log\left(  1-\alpha\right)  ~,
\end{equation}
where the coefficients $c_{1}=\frac{1}{4}\pi+\frac{1}{64}\sqrt{2}%
\simeq0.807\,50$, $c_{2}=-\frac{1}{128}\left(  \pi+3\sqrt{2}\right)
\simeq-5.768\,9\times10^{-2}$, $c_{3}=\frac{49}{30}\sqrt{2}-\frac{47}{64}%
\pi\simeq2.775\times10^{-3}$, $c_{4}=\left(
\frac{1}{2}-\frac{1}{128}\right) \pi-\left(
\frac{11}{10}-\frac{1}{128}\right) \sqrt{2}\simeq1.666\,3\times
10^{-3}$, $\check{c}=\frac{1}{64}\sqrt{2}\simeq2.209\,7\times10^{-2}$
are chosen so as to reproduce the calculated properties.  Numerical
integration of Eq.~(\ref{A h}) (evaluation of the integrand at large
$q$ requires some care) shows that the maximum error in this
approximation is less than $5\times 10^{-5}$.  As the cutoff scale is
large, $\bar{x}$ and $\bar{y}$ typically are small, and such attention
to the small--$\bar{x}$--and--$\bar{y}$ limit is appropriate.


\begin{thebibliography}{99}
\bibitem {vdw}For alkali atoms, the intermolecular potential scales
asymptotically as $V(r) = - C_{6}/r^{6}$, and the range of the
interaction $r_{0}$ can be taken as the van der Waals length, $r_{0} =
(m C_{6}/\hbar^{2})^{1/4}$.  The condition then reads $n^{-1/3} \gg
r_{0}$.
\bibitem {Hutchinson97}D. A. W. Hutchinson, E. Zaremba and A. Griffin, 
Phys. Rev. Lett. \textbf{78}, 1842 (1997).
The  equation for the condensate here is 
the same as that in \cite{MCT97}, but the treatment of the thermal 
cloud is more advanced.
\end{thebibliography}
\end{document}